\documentclass[10pt,journal,compsoc]{IEEEtran}
\usepackage{amsthm,amsmath,amsfonts,amssymb,bm}
\usepackage{array}
\usepackage{textcomp}
\usepackage{stfloats}
\usepackage{url}
\usepackage{verbatim}
\usepackage{cite}
\usepackage{dsfont}
\usepackage{enumitem}
\usepackage{multirow}
\usepackage{graphicx}
\usepackage{subcaption}
\usepackage{caption}
\usepackage{threeparttable}
\usepackage[table,xcdraw]{xcolor}
\usepackage{bbding}

\allowdisplaybreaks[4]

\usepackage{setspace} 

\usepackage[noend]{algpseudocode}
\usepackage{algorithmicx,algorithm}

\renewcommand{\arraystretch}{0.90}

\graphicspath{{figures/},{figures/photos/},{figures/experiments/}\:}
\DeclareGraphicsExtensions{.pdf,.jpg,.png}

\floatstyle{boxed}
\newfloat{fig}{bt}{fig}
\floatname{fig}{Fig.}
\AtBeginDocument{\captionsetup[fig]{labelsep=period}}
\newtheoremstyle{tightdef}  
  {2pt}   
  {2pt}   
  {\normalfont} 
  {}      
  {\bfseries} 
  {.}     
  {.5em}  
  {}      

\theoremstyle{tightdef}
\newtheorem{definition}{Definition}
\newtheorem{theorem}{Theorem}

\newtheorem{corollary}{Corollary}
 \newtheorem{remark}{Remark}

\usepackage{xspace}

\begin{document}

\title{Secure Multi-Key Homomorphic Encryption with Application to Privacy-Preserving Federated Learning}
\author{Jiahui~Wu, Tiecheng Sun,~\IEEEmembership{Member,~IEEE,} Fucai~Luo, Haiyan~Wang, Weizhe~Zhang,~\IEEEmembership{Senior Member,~IEEE}
\thanks{This work was supported in part by the National Key Research and Development Program of China (Grant No. 2025YFE0200100), and in part by the Key Program of the Joint Fund of the National Natural Science Foundation of China (Grant No.U22A2036).}
\thanks{Jiahui Wu, Tiecheng Sun, Haiyan Wang, and Weizhe Zhang are with Pengcheng Laboratory, Shenzhen 518000, China;
Fucai Luo is with the School of Computer Science and Technology, Zhejiang Gongshang University, Hangzhou, China
(e-mail: wujh01@pcl.ac.cn; tiechengsun@126.com; lfucai@126.com; wanghy01@pcl.ac.cn; wzzhang@hit.edu.cn.);
Corresponding author: Weizhe Zhang.}}

\maketitle

\begin{abstract}
Multi-Key Homomorphic Encryption (MKHE), proposed by L\'{o}pez-Alt et al. (STOC 2012), allows for performing arithmetic computations directly on ciphertexts encrypted under distinct keys. Subsequent works by Chen and Dai et al. (CCS 2019) and Kim and Song et al. (CCS 2023) extended this concept by proposing multi-key BFV/CKKS variants, referred to as the CDKS scheme. These variants incorporate asymptotically optimal techniques to facilitate secure computation across multiple data providers.
In this paper, we identify a critical security vulnerability in the CDKS scheme when applied to multiparty secure computation  tasks, such as privacy-preserving federated learning (PPFL).
In particular, we show that CDKS may inadvertently leak plaintext information from one party to others.
To address this issue, we propose a new scheme, SMHE (Secure Multi-Key Homomorphic Encryption), which incorporates a novel masking mechanism into the multi-key BFV and CKKS frameworks to ensure that plaintexts remain confidential throughout the computation.
We implement a PPFL application using SMHE and demonstrate that it provides significantly improved security with only a modest overhead in homomorphic evaluation.
The code is publicly available at \url{https://github.com/JiahuiWu2022/SMHE.git}.
\end{abstract}

\begin{IEEEkeywords}
Multi-key homomorphic encryption, masking scheme, privacy protection, federated learning.
\end{IEEEkeywords}


\section{Introduction}

Homomorphic encryption (HE) is a cryptographic technique that enables computations to be performed directly on encrypted data, eliminating the need for decryption during the process. This property allows for the secure processing of sensitive data while preserving its confidentiality.
For decades, constructing a HE scheme, 
capable of supporting arbitrary computations on ciphertexts, remained an open problem until Gentry's groundbreaking work \cite{gentry2009fully}. Since then, significant advancements have been achieved in the field, leading to various HE schemes such as BFV \cite{brakerski2012fully,fan2012somewhat}, GSW \cite{gentry2013homomorphic}, BGV \cite{brakerski2014leveled}, TFHE \cite{chillotti2016faster}, and CKKS \cite{cheon2017homomorphic}. These schemes have expanded the practicality and efficiency of HE, making it a vital tool for secure computations in modern applications.
One of the key features of HE is its ability to enable secure computation ``on-the-fly," meaning that data owners are not required to be actively involved during the computation process. Instead, the evaluation can be carried out entirely by a public server. This makes HE particularly appealing for scenarios such as cloud computing, where data is processed remotely, and privacy preservation is of paramount importance.

In recent years, the demand for secure multiparty computation (MPC) protocols has surged, driven by applications such as federated learning (FL) \cite{mcmahan2021advances}. MPC allows multiple parties to collaboratively evaluate a function or circuit on their private inputs without revealing any information beyond the final result. However, traditional single-key HE schemes are not well-suited for such multi-party settings.
A major limitation arises when multiple data sources are involved, as standard HE schemes typically require all data to be encrypted under the same encryption key.
This requirement grants the entity possessing the corresponding decryption key full access to the encrypted data, which raises issues regarding privacy and the reliance on centralized trust.

To address these limitations, researchers have extended the functionality of HE through approaches such as threshold homomorphic encryption (THE) \cite{asharov2012multiparty,boneh2018threshold,mouchet2023efficient,mouchet2021multiparty,park2021homomorphic,ma2022mkhe} and multi-key homomorphic encryption (MKHE) \cite{lopez2012fly,clear2015multi,mukherjee2016two,peikert2016multi,chen2019multi,ccs2019mkhe,ccs2023mkhe}. THE allows decryption authority to be distributed among multiple parties, ensuring that no single entity possesses full access to the secret key. Similarly, MKHE enables computations on ciphertexts encrypted under different keys, thereby avoiding the need for a single shared encryption key and mitigating the risk of authority concentration.
These extensions not only overcome the limitations of single-key HE but also integrate seamlessly with secure MPC protocols, preserving the inherent advantages of HE. Recognizing their potential, the National Institute of Standards and Technology (NIST) has highlighted these primitives as promising candidates for standardization in its recent call for multi-party threshold cryptographic schemes \cite{brandao2023nist}. As a result, THE and MKHE have emerged as essential building blocks for advancing privacy-preserving technologies in collaborative and distributed environments.

In THE, multiple parties collaboratively generate a common public key, with the corresponding secret key shared among them through a secret-sharing mechanism. Although THE schemes generally achieve performance levels comparable to single-key HE and are typically more efficient than multi-key HE schemes, they suffer from a critical limitation: reliance on a static key access structure. In other words, all participants must be predetermined and fixed during the initial setup phase.
In contrast, this work focuses on MKHE, which offers significant advantages in terms of flexibility and reduced interaction requirements. Specifically, MKHE schemes allow each participant to independently generate their own public-secret key pairs without needing knowledge of other parties' keys. This property enables operations on ciphertexts encrypted under different keys and allows computations to be performed in a public cloud environment without establishing a common public key. Moreover, recent advances in MKHE have introduced \textit{full dynamism}, allowing computations to be performed on multi-key ciphertexts without predefined circuits. Arbitrary circuits can be evaluated in real time, and new participants or ciphertexts can be incorporated into ongoing evaluations at any stage. This flexibility facilitates the construction of SMC protocols on top of MKHE, leveraging its dynamic and adaptive nature \cite{mukherjee2016two}. The ability to seamlessly integrate new participants and ciphertexts on-the-fly is a key advantage, making MKHE particularly well-suited for applications such as FL, where the set of participants may change throughout the computation process.

While MKHE provides a flexible and dynamic framework, designing secure and efficient MKHE schemes is significantly more challenging than for other homomorphic encryption variants, due to stringent functional and security requirements. Since the pioneering work of L\'{o}pez-Alt et al. \cite{lopez2012fly}, who introduced the first MKHE scheme based on NTRU, considerable efforts have been made to extend traditional (single-key) HE schemes into their multi-key counterparts \cite{clear2015multi,brakerski2016lattice,mukherjee2016two,peikert2016multi,chen2017batched,chen2019multi,ccs2019mkhe,ccs2023mkhe}. Nonetheless, achieving both strong security guarantees and practical efficiency in MKHE remains a substantial challenge. Early MKHE constructions often suffered from high computational and communication overhead, limiting their practicality in real-world deployments. Recent advancements \cite{chen2019multi,ccs2019mkhe,ccs2023mkhe} have proposed improved designs with significantly enhanced asymptotic and concrete efficiency, representing the current state-of-the-art in MKHE. However, despite these improvements, previously overlooked security vulnerabilities have emerged, making these MKHE schemes potentially less secure than their single-key HE counterparts.

\subsection{Challenges and Contributions}\label{ssec:contributions}
Building on the above observation, we revisit the security assumptions behind CDKS-style MKHE and make the challenges and our remedies explicit. We keep technical details lightweight here and defer full derivations to later sections.

\textbf{Challenge 1: CDKS violates the only-the-result goal in MPC.}
In the CDKS expansion + distributed-decryption workflow, the expanded component for party $i$ is tied to that party's fresh share, and the merge uses the observable pair $(c_{i0},\nu_i)$ (here $ct_i=(c_{i0},c_{i1})$ denotes a fresh BFV/CKKS ciphertext under key $i$, and $\nu_i$ is party $i$'s partial decryption share).
Consequently, any party that sees $(c_{i0},\nu_i)$ can recover client $i$'s plaintext via the one-line attack
\[
\Big\lfloor \tfrac{t}{Q}\big(c_{i0}+\nu_i\big)\Big\rceil \approx \mu_i \quad \text{(cf.\ Eq.~\eqref{eq:inference})},
\]
contradicting the MPC ideal functionality that only the round result be revealed.

\textbf{Remedy (structured masking).}
To eliminate this leakage channel, we attach to each input a \emph{structured masking} triple $(ct,cz,\Gamma)$ and derive an auxiliary ciphertext $cx$ so that all intermediate states are statistically independent and become informative only at the final linear merge. Our design is tailored to CKKS/BFV and remains compatible with multi-key evaluation. Conceptually, masking is an encryption of fresh randomness that is added to the ciphertext; because the mask is itself a semantically secure ciphertext, intermediate views reveal nothing about the underlying message.

\noindent\textbf{Provenance.}
The masking paradigm we adopt is \emph{not ad hoc}: it adapts masking primitives \emph{originally developed} for GSW-based multi-key HE (MKGSW) \cite{clear2015multi, mukherjee2016two} to the polynomial-ring setting of BFV/CKKS via a new ciphertext expansion and evaluation procedure. This preserves the only-the-result goal by removing the leakage path through $(c_{i0},\nu_i)$ while sustaining efficient multi-key computation.

\textbf{Challenge 2: Why MKGSW-style masking does not drop in for CKKS/BFV.}
While the MKGSW masking scheme ensures security and correctness in the GSW matrix setting, a direct transplant to CKKS/BFV fails for two reasons:

\noindent\textbf{R1 (structural mismatch).}
GSW ciphertexts are matrices $C\in\mathbb{Z}_Q^{n\times m}$, whereas CKKS/BFV ciphertexts are polynomial pairs $ct=(c_0,c_1)\in R_Q^2$. This makes it nontrivial to realize a CKKS/BFV joint decryption relation of the form
\[
\langle sk, (ct+cx) \rangle + \langle sk', ct \rangle \approx \mu\cdot \theta,
\]
that mirrors the MKGSW equation $\big(sk(C+X)+sk'C\big)\cdot \Theta^{-1}W \approx \mu$ (here $cx$ would be a mask derived from auxiliary info $\Gamma$, and $\theta$ and $\Theta$ are two fixed term).

\noindent\textbf{R2 (quadratic blow-up).}
The GSW masking scheme requires large extended ciphertexts for correct demasking, with each single-bit encryption resulting in an $O(n^2)$-size expanded ciphertext in the number of parties\footnote{In the $n$-party setting, expansion realizes an $n\times n$ block layout: diagonal blocks carry per-party data while off-diagonal positions provide masking hooks to enable joint demasking. Both the row and column dimensions grow linearly with $n$, so the encoded length scales as $O(n^2)$ times a base GSW block, even when many entries are structurally zero.}.
This leads to significant computational and communication overhead.
Therefore, it is necessary to design a new masking scheme for CKKS/BFV that ensures the expanded ciphertext size scales linearly with $n$, thereby improving both computational and space efficiency.

\textbf{Remedy (linear-size masked expansion).}
We propose a CKKS/BFV-specific masking scheme that achieves a linear-size expansion (i.e., the \emph{expanded} ciphertext size scales \emph{linearly} in $n$) and a correct joint decryption. In addition to a masking ciphertext $cx=(x_0,x_1)$ (carrying a random mask $r$), we generate a ``demasking'' ciphertext $cz=(z_0,z_1)$ and enforce
\begin{equation}\label{eq:construction}
  \langle sk, cx \rangle + \langle sk', cz \rangle \approx 0.
\end{equation}
Then the joint decryption of the \emph{expanded} ciphertext
\[
\overline{ct} = (z_0 + x_0 + c_0,\; x_1 + c_1,\; z_1) \triangleq (\bar{c}_0,\bar{c}_1,\bar{c}_2)
\]
under the concatenated key $\overline{sk}=(1,s,s')$ yields
\[
\langle \overline{sk}, \overline{ct} \rangle
= \langle sk, ct \rangle + \underbrace{\big(\langle sk, cx \rangle + \langle sk', cz \rangle\big)}_{\approx 0}
\approx \mu,
\]
and the size of $\overline{ct}$ is linear in the number of participants.

\noindent\textit{Instantiation (CKKS; BFV is analogous).}
Let $ct\leftarrow\texttt{Encrypt}(pk,\mu)=w\cdot pk+(\mu+e_0,e_1)\bmod Q$, where $sk=(1,s)$, $pk=(b=-as+e,\,a)$, and $e,e_0,e_1$ are small. Choose $cz\leftarrow\texttt{Encrypt}(pk,0)=r\cdot pk+(e_{r_0},e_{r_1})$ so that
\[
\langle sk', cz \rangle = r(b-b') + e_z \bmod Q,\quad b'=-as'+e' \bmod Q.
\]
Encrypt the mask $r$ as auxiliary information $\Gamma\leftarrow\texttt{Encrypt}(pk,r)=w\cdot pk+(r+e_{x_0},e_{x_1})$, and set
$cx \;=\; \Gamma\,(b'-b),$
so that, within the decryption noise margin, $\langle sk, cx\rangle$ cancels $\langle sk', cz\rangle$ and thereby satisfies Eq.~\eqref{eq:construction}.

\textit{$n$-party generalization (linear growth).}
For $n$ parties with secret keys $sk_i=(1,s_i)$, we create per-party masked pairs $(cx_i,cz_i)$ and aggregate them into a single expanded ciphertext whose coordinate-length increases by exactly one for each additional party:
\[
\overline{ct}=(\bar{c}_0,\bar{c}_1,\ldots,\bar{c}_n),
\qquad
\overline{sk}=(1,s_1,\ldots,s_n).
\]
The per-party cancellation constraints (generalizing Eq.~\eqref{eq:construction}) ensure that mask contributions vanish in the final inner product, giving the unified joint decryption
\begin{equation}\label{eq:DecryptExpandedCiphertext}
\langle \overline{sk}, \overline{ct} \rangle
= \big\langle (1,s_1,\ldots,s_n),\,(\bar{c}_0,\ldots,\bar{c}_n)\big\rangle
\approx \mu.
\end{equation}
Hence the expanded ciphertext has $(n{+}1)$ coordinates and its size scales \emph{linearly} in $n$. 

\textbf{Challenge 3: Noise growth breaks naive demasking.}
The naive choice $cx=\Gamma\,(b'-b)$ introduces a large error term due to multiplication by the essentially random polynomial $(b'-b)\in R_Q$:
\begin{align*}
\langle sk, cx \rangle
&= \langle sk, (b'-b)\,(w\cdot pk + (r+e_{x_0},e_{x_1})) \rangle \\
&= (b'-b)\big((w\cdot b + r + e_{x_0}) + (w\cdot a + e_{x_1})s\big) \\
&= r(b'-b) + \underbrace{(b'-b)(we + e_{x_0} + e_{x_1}s)}_{E} \;\bmod\ Q,
\end{align*}
where $E\in R_Q$ can be non-negligible. Consequently, the desired cancellation
$\langle sk, cx \rangle + \langle sk', cz \rangle \approx 0$
fails, invalidating the decryption equation.

\textbf{Remedy (Gadget decomposition to tame noise).}
We control the error growth by applying gadget decomposition to the factors that amplify noise, reducing $E$ so that
$\langle sk, cx \rangle + \langle sk', cz \rangle \approx 0$ holds with the necessary tightness. Intuitively, gadget techniques linearize large multipliers into well-behaved digit expansions that keep the induced error within the decryption margin. We provide the decomposition details in Section~\ref{ssec:gedget} and the full masking construction in Section~\ref{ssec:maskingScheme}.

\section{Related Work}\label{sec:relatedWork}

MKHE enables computations on ciphertexts encrypted under distinct keys.
L\'{o}pez-Alt and Wichs \cite{lopez2012fly} first introduced an MKHE scheme based on the NTRU cryptosystem,
whose security relies on a relatively non-standard assumption concerning polynomial rings. This assumption differs from the more widely used Learning With Errors (LWE) assumption \cite{regev2009lattices} or its ring-based variant \cite{lyubashevsky2013ideal}, and it currently lacks a worst-case hardness theorem \cite{peikert2016multi} to support its security.
Subsequent work by Clear et al. \cite{clear2015multi} proposed an LWE-based MKHE scheme that employs a multikey variant of the GSW scheme and ciphertext extension techniques. Mukherjee et al. \cite{mukherjee2016two} later simplified this approach. Both of these schemes design masking scheme for the GSW cryptosystem. Specifically, the masking system enables evaluators to generate a specific mask by combining a universal mask and a target participant identity. This mask facilitates the ``encoding" of ciphertexts from different identities into a larger matrix, which allows for joint homomorphic computations on these ciphertexts.
Mukherjee et al. also proposed a $2$-round (plain) MPC protocol in the common random string (CRS) model for secure distributed decryption. Peikert and Shiehian \cite{peikert2016multi} further extended the multi-key GSW scheme to develop two multi-hop MKHEs. However, all these GSW-based MKHE variants face a major limitation: they can only encrypt a single bit in a large expanded GSW ciphertext, which results in substantial space and time complexities as the bit-length of the ciphertext grows quadratically with $N^2$, where $N$ is the number of distinct participants.
Brakerski et al. \cite{brakerski2016lattice} proposed an MKHE scheme based on LWE that employs short ciphertexts and a quasi-linear expansion rate. However, both the asymptotic and concrete efficiency of this scheme have not been clearly understood, making its practical applicability uncertain.

A follow-up study by Chen et al. \cite{chen2019multi} introduced a multi-key TFHE. 
Meanwhile, another line of research \cite{ccs2019mkhe,ccs2023mkhe} has focused on designing multi-key variants of batch HEs, such as BFV \cite{brakerski2012fully,fan2012somewhat} and CKKS \cite{cheon2017homomorphic}. However, these constructions exhibit significant security vulnerabilities and thus MPC applications such as PPFL \cite{tifs2024application} and secure distributed sparse Gaussian processes \cite{aaai2024application} that utilize the multi-key CKKS scheme fail to meet their stated security objectives.

In this paper, we propose new multi-key variants of batch HE schemes, including BFV and CKKS, by introducing a novel masking scheme for the CKKS/BFV cryptosystem. Our scheme supports the encryption of both floating-point and integer values with multiple bits, without requiring the large ciphertext expansion as in \cite{clear2015multi,mukherjee2016two,peikert2016multi}. Furthermore, our scheme addresses the security vulnerabilities present in existing multi-key BFV/CKKS constructions \cite{ccs2019mkhe,ccs2023mkhe}.


\section{Preliminaries}\label{sec:preliminaries}

In this section, we present the necessary background for this work. A summary of the mathematical notations used throughout the paper is provided in Table~\ref{tab:notations}.

\begin{table}
 \footnotesize
 \centering
 \renewcommand\arraystretch{1.23}
 \caption{The symbols and their corresponding interpretations.}\label{tab:notations}
 \begin{tabular}{m{1.5cm}|m{6.5cm}}
  \hline
  \textbf{Symbols}   & \textbf{Interpretations}\\\hline \hline
  $n$ $/$ $N$& The number of the clients $/$ The RLWE dimension\\
  \hline
  $\mathcal{H}$ $/$ $\lfloor \cdot \rceil$& Gadget decomposition function $/$ The rounding function\\
  \hline
  $\mathbf{g}$, $\tau$ & Gadget vector and its dimension \\
  \hline
  $\lambda$ & Security parameter\\
  \hline
  $\mathbb{Z}_Q$ & $\mathbb{Z}\cap (-Q/2,Q/2]$\\
  \hline
  $\langle u,v\rangle$ & The inner product of two tuples/vectors $u,v$\\
  \hline
  $||a||_{\infty}$ & The $\ell^{\infty}$-norm of the coefficient vector of $a$\\
  \hline
  $\chi$ $/$ $U(\cdot)$ & $\chi$ distributions over $R$ / Uniform distribution\\
  \hline
  $D_{\sigma}$ & Discrete Gaussian distribution with standard deviation $\sigma$\\
  \hline
  $a\leftarrow A$ & $a$ is sampled from a set or distribution $A$ \\
  \hline
  $ct$ $/$ $\overline{ct}$ $/$ $\widehat{ct}$ & Fresh ciphertext/Expanded ciphertext/Masked ciphertext\\
  \hline
  $c_j^i$ $/$ $\bar{c}_j^i$ & The $j$th component of the ciphertext $ct_i$ $/$ $\overline{ct}_i$\\
  \hline
 \end{tabular}
\end{table}

\subsection{Ring Learning with Errors}\label{ssec:Prelim_RLWE}

The Ring Learning With Errors (RLWE) assumption is based on polynomial arithmetic with coefficients in a finite field.
Specifically, let $N$ be a power of two. Define \( R = \mathds{Z}[x]/(x^N + 1) \) and \( R_Q = \mathbb{Z}_Q[x]/(x^N + 1) \).
Let \( \chi \) be a probability distribution over \( R \), and \( \sigma > 0 \) be a real number. The RLWE assumption, defined by the parameters \( (N, Q, \chi, \sigma) \), asserts that it is computationally infeasible to distinguish between two scenarios: given polynomially many samples of either \( (b, a)\in R_Q^2 \) or \( (a \cdot s + e,a)\in R_Q^2 \), where \( s \) is drawn from \( \chi \) and \( e \) is sampled from the discrete Gaussian distribution \( D_\sigma \) with mean 0 and standard deviation $\sigma$, the distribution of the two cases remains indistinguishable.
The RLWE assumption serves as the foundation for homomorphic encryption schemes like BFV and CKKS.

\vspace{-10pt}

\subsection{Multi-Key Homomorphic Encryption}\label{ssec:mkhe}

Multi-Key homomorphic encryption (MKHE) is an cryptographic scheme that extends the capabilities of traditional single-key HE to support computations on ciphertexts encrypted using distinct keys.
It contains the following algorithms:

\begin{itemize}[leftmargin=10pt,itemsep=0pt, topsep=0pt, parsep=0pt, partopsep=0pt]
  \item $pp\leftarrow\texttt{MKHE.Setup}(1^{\lambda})$. Generates a set of public parameters $pp$ based on a given security parameter $\lambda$.

  \item $(sk,pk)\leftarrow\texttt{MKHE.KeyGen}(pp)$. Generates a secret-public key pair $(sk,pk)$ using the public parameters $pp$.

  \item $ct\leftarrow\texttt{MKHE.Encrypt}(\mu,pk)$. Encrypts the plaintext $\mu$ with the public key $pk$ and outputs the fresh ciphertext $ct$.

  \item $\overline{ct}\leftarrow\texttt{MKHE.Expand}\big(\{pk_{i}\}_{i\in [1,n]}, i,ct\big)$. Expands the given ciphertext $ct$, encrypted under $pk_i$, into an expanded ciphertext $\overline{ct}$ associated with $n$ public keys $\{pk_{i}\}_{i\in [1,n]}$.

  \item $\overline{ct}\leftarrow\texttt{MKHE.Eval}(\mathcal{C};\overline{ct}_1,\cdots,\overline{ct}_k;\{pk_{i}\}_{i\in [1,n]})$.
    Performs homomorphic evaluation on the circuit $\mathcal{C}$ over the ciphertexts $\overline{ct}_1,\cdots,\overline{ct}_k$ associated with the joint public key set $\{pk_{i}\}_{i\in [1,n]}$
    and outputs the evaluated ciphertext $\overline{ct}$. 

  \item $\mu:=\texttt{MKHE.Decrypt}(\overline{ct};\{sk_{i}\}_{i\in [1,n]})$.  Recovers the plaintext $\mu$ using the corresponding private keys $\{sk_{i}\}_{i\in [1,n]}$ of all public keys referenced in the given ciphertext $\overline{ct}$.
\end{itemize}

A unique feature of MKHE is its use of ``reference sets" $\{pk_{i}\}_{i\in [1,n]}$. Each expanded ciphertext maintains a reference to the public keys under which the data has been encrypted. Initially, a fresh ciphertext is tied to a single key. As homomorphic computations progress and involve ciphertexts encrypted under additional keys, the reference set expands.
Decryption requires all secret keys corresponding to the keys in the reference set.
Specifically, in collaborative scenarios, each party partially decrypts the ciphertext with his secret key and broadcasts the partial decryption result.
Then, the plaintext can be constructed by combining all parties' partial decryption results. This distributed decryption is as follows: 

\begin{itemize}[leftmargin=10pt,itemsep=0pt, topsep=0pt, parsep=0pt, partopsep=0pt]
  \item 
  $\nu_{i}:=\texttt{MKHE.PartDec}(\overline{ct},sk_{i})$.  Partially decrypts the ciphertext $\overline{ct}$ using the secret key $sk_{i}$ and returns the partial decryption result $\nu_{i}$.

  \item 
  $\mu:=\texttt{MKHE.FullDec}(\overline{ct},\{\nu_{i}\}_{i\in [1,n]})$.  Fully decrypts to obtain the plaintext $\mu$ by combining all partial decryption results $\{\nu_{i}\}_{i\in [1,n]}$ from parties in the reference set of $\overline{ct}$.
\end{itemize}

The correctness and security of MKHE is defined as follows.
\begin{definition}[Correctness]
    Let $\overline{ct}_1, \cdots, \overline{ct}_k$ be MKHE ciphertexts encrypting messages $\mu_1,\cdots,\mu_k$, respectively. Denote by $\{pk_i\}_{i\in [1,n]}$ the reference set associated with these ciphertexts. Suppose $\overline{ct} \leftarrow $ $\texttt{MKHE.Eval}(\mathcal{C}; \overline{ct}_1, \cdots, \overline{ct}_k; \{pk_i\}_{i\in [1,n]})$ is the result of evaluating a circuit $\mathcal{C}$ over these ciphertexts. Then, decryption using the corresponding secret keys $\{sk_i\}_{i\in [1,n]}$ correctly recovers the result with overwhelming probability:
    $$
    \texttt{MKHE.Dec}(\overline{ct}; \{sk_i\}_{i\in [1,n]}) = \mathcal{C}(\mu_1, \cdots,\mu_k).
    $$
    For approximate encryption schemes such as CKKS, this correctness notion is relaxed to allow for small errors: 
    $$\texttt{MKHE.Dec}(\overline{ct}; \{sk_i\}_{i\in [1,n]}) \approx \mathcal{C}(\mu_1, \cdots,\mu_k).$$
\end{definition}

\begin{definition}[Simulation-Based Security]
\label{definition:simulationSecurity}
An MKHE scheme is secure if it is \emph{simulation-based secure} that any real-world adversary interacting with the system 
cannot learn more than what is revealed by an ideal functionality.

Formally, let $\lambda$ be the security parameter. Consider a set of $n$ parties ($n \geq 2$), each holding a message $\mu_i$ and a key pair $(pk_i, sk_i) \leftarrow \texttt{MKHE.KeyGen}(pp)$, where $pp \leftarrow \texttt{MKHE.Setup}(1^\lambda)$. The parties perform the following:

\begin{itemize}[leftmargin=10pt, itemsep=2pt, topsep=0pt, parsep=0pt, partopsep=0pt]
  \item Encryption: $ct_i \leftarrow \texttt{MKHE.Encrypt}(pk_i, \mu_i)$;
  \item Expansion: $\overline{ct}_i\leftarrow\texttt{MKHE.Expand}\big(\{pk_i\}_{i\in [1,n]},i,ct_i\big)$;
  \item Evaluation: $\overline{ct} \leftarrow \texttt{MKHE.Eval}(\mathcal{C}; \{ \overline{ct}_i \}_{i \in [1,k]}; \{ pk_i \}_{i \in [1,n]})$; 
  \item Decryption: $\nu_{i}:=\texttt{MKHE.PartDec}(\overline{ct},sk_{i})$ and $\mu:=\texttt{MKHE.FullDec}(\overline{ct},\{\nu_{i}\}_{i\in [1,k]})$, where $2\leq k\leq n$.
\end{itemize}

Let $\mathcal{A}$ be a real-world adversary who observes all ciphertexts, public keys, partial decryption, and the evaluation result. Then there exists a probabilistic polynomial-time (PPT) simulator $\mathsf{Sim}$ such that the following distributions are computationally indistinguishable (denoted by $\approx_c$):

\begin{itemize}[leftmargin=10pt, itemsep=2pt, topsep=0pt, parsep=0pt, partopsep=0pt]
  \item \textbf{Real-world view:}
  The adversary's view in the real protocol execution, including the public parameters $pp$, public keys $\{pk_i\}$, input ciphertexts $\{ct_i,\overline{ct}_i\}$, evaluated ciphertext $\overline{ct}$, partial decryption $\{\nu_i\}$, and optionally the final output $\mu$.

  \item \textbf{Ideal-world simulation:}
  The simulated view produced by $\mathsf{Sim}$, given only the public parameters $pp$ and the final output $\mu = \mathcal{C}(\mu_1, \dots, \mu_k)$. 
\end{itemize}
We say the scheme is secure if:
$\text{View}_{\mathcal{A}}^{\text{real}}(\lambda) \approx_c \text{View}_{\mathsf{Sim}}^{\text{ideal}}(\lambda),$
i.e., no efficient adversary can distinguish between the real view and the ideal simulation with non-negligible probability.
\end{definition}

\subsection{Gadget Decomposition}\label{ssec:gedget}
In lattice-based HE schemes, the accumulation of noise during homomorphic operations poses a significant challenge to the efficiency and correctness of computations. One widely used technique for mitigating this noise growth is \textit{Gadget Decomposition}.
By leveraging structured representations, it provides a mechanism to control noise in ciphertexts, enabling accurate homomorphic computations.
Informally, gadget decomposition is to represent elements in a ring as compact, structured linear combinations of predefined basis elements.

\begin{definition}[Gadget Decomposition]
   Let $Q$ and $\tau$ be a modulus and a positive integer, respectively.
   A gadget decomposition is defined as a mapping $\mathcal{H}: R_Q \to R^{\tau} $ that meets the conditions below for all $b \in R_Q$:

  \begin{itemize}[leftmargin=10pt, itemsep=2pt, topsep=0pt, parsep=0pt, partopsep=0pt]
    \item \textbf{Reconstruction property:}
    A constant vector $\bm{g} = (g_0, g_1,$ $ \cdots, g_{\tau-1}) \in R_Q^{\tau}$ exists such that $\langle \mathcal{H}(b), \bm{g} \rangle \equiv b \pmod{Q}$.
    \item \textbf{Bounded coefficients:} The coefficients of \( \mathcal{H}(b) \) are small, i.e., \( \| \mathcal{H}(b) \|_\infty \leq B_\mathcal{H} \) for some constant \( B_\mathcal{H} > 0 \).
  \end{itemize}
\end{definition}

The vector $\bm{g}$ is referred to as the gadget vector, while \( \mathcal{H}(b) \) is a compact representation of $b$ with bounded coefficients. The mapping \( \mathcal{H} \) can be viewed as a right inverse of the inner product operation \( \mathcal{G}(\bm{u}) = \langle \bm{g},\bm{u} \rangle \pmod{Q} \), ensuring that \( \mathcal{G}(\mathcal{H}(b)) = b \).
\begin{definition}[Gadget Encryption]
  For a given message $\mu\in R$ and a secret key $sk=(1,s)$ with $s\in R$, a pair $\bm{\Gamma}=(\bm{\varsigma}_0,\bm{\varsigma}_1)\leftarrow\texttt{GgtEnc}(sk,\mu)\in R_Q^{\tau\times 2}$ is defined as a gadget encryption
  of the message if its decryption meets $\langle sk,\bm{\Gamma}\rangle\approx \mu\cdot\bm{g} \pmod{Q}$.
\end{definition}
\begin{definition}[External Product]
    Define $b \boxdot \bm{\varsigma}=\langle \mathcal{H}(b),\bm{\varsigma}\rangle \pmod Q$ as the external product of $b$ and $\bm{\varsigma}$, where $b\in R_Q$ and $\bm{\varsigma}\in R_Q^{\tau}$.
    Additionally, for $\bm{\Gamma}=(\bm{\varsigma}_0,\bm{\varsigma}_1)\in R_Q^{\tau\times 2}$, we define $b \boxdot \bm{\Gamma}=(b \boxdot \bm{\varsigma}_0,b\boxdot \bm{\varsigma}_1)$.
\end{definition}
By employing the gadget decomposition technique, it becomes feasible to homomorphically perform multiplication on arbitrary ring elements while avoiding the generation of excessive noise. Specifically, let $\bm{\Gamma} = (\bm{\varsigma}_0, \bm{\varsigma}_1) \in R_Q^{\tau\times 2} $ represent a gadget encryption of \( \mu \in R \) using \( sk \), satisfying
$\langle sk,\bm{\Gamma}\rangle= \mu\cdot\bm{g}+\bm{e}\ (\mathrm{mod} \, Q)$ for a small \( \bm{e} \in R^{\tau} \). The external product \( (c_0, c_1) \leftarrow b\, \boxdot \, \bm{\Gamma} \) then meets:
\begin{equation*}\label{eq:externalProduct}
    \begin{split}
      \langle sk,(c_0,c_1)\rangle & =\langle sk,(b\boxdot\bm{\varsigma}_0,b\boxdot\bm{\varsigma}_1)\rangle
      = \langle \mathcal{H}(b),\langle sk,\bm{\Gamma}\rangle\rangle \\
        & = \langle \mathcal{H}(b),\mu\cdot\bm{g}+\bm{e}\rangle=b\cdot\mu+e\ (\mathrm{mod} \, Q),
    \end{split}
\end{equation*}
where the noise term $e = \langle \mathcal{H}(b), \bm{e}\rangle \in R$ remains small. Consequently, $ (c_0, c_1) $ can be viewed as a noisy encryption of $b \cdot \mu $, as intended.

\section{Overview of Prior Work}\label{sec:priorWork}

This section reviews the most relevant researches proposed by Chen and Dai et al. \cite{ccs2019mkhe} and Kim and Song et al. \cite{ccs2023mkhe}.
The former designs multi-key BFV/CKKS variants, while the latter focuses on enhancing computational efficiency and mitigating noise growth in these schemes.
Collectively, we refer to these schemes as CDKS. We begin by outlining the core construction of CDKS and then highlight a critical security vulnerability within the framework.

\subsection{Foundational Construction of CDKS}

CDKS is built on the CRS model, where all key holders share access to identical public random polynomials. Specifically, tt contains the following algorithms.
Components specific to the \textcolor{blue}{multi-key CKKS} and \textcolor{red}{multi-key BFV} schemes are highlighted in blue and red, respectively.

\begin{itemize}[leftmargin=10pt,itemsep=0pt, topsep=0pt, parsep=0pt, partopsep=0pt]
  \item $pp\leftarrow\texttt{CDKS.Setup}(1^{\lambda})$:
  Let \( N \) denote the RLWE dimension, \( Q = \prod_{i=0}^{L} q_i \) represent the ciphertext modulus for some integers \( q_i \), the plaintext modulus \( t \in \mathbb{Z} \), the key distribution \( \chi \) over \( R \), and the error distribution \( D_{\sigma} \) with $\sigma$ be positive value.
  $\mathbf{a}\leftarrow U(R_Q^k)$ is a CRS.
  and the public parameter is defined as \( pp = (N, \textcolor{red}{t}, Q, \chi, \sigma, \mathbf{a}) \).

  \item $(sk,pk,evk)\leftarrow\texttt{CDKS.KeyGen}(pp)$: Generates secret and public keys $(sk=(1,s),pk=(b,a))$ and an evaluation key $evk=(\mathbf{b},\mathbf{d},\mathbf{u},\mathbf{v})$:
    Sample $s,\gamma\leftarrow\chi$, $\mathbf{e}_0,\mathbf{e}_1,\mathbf{e}_2\leftarrow D_{\sigma}^{\tau}$, $\mathbf{u} \leftarrow U(R_Q^{\tau})$
     and compute $\mathbf{b}=-s\cdot\mathbf{a}+\mathbf{e}_0 \pmod Q$, $\mathbf{d} = -\gamma \cdot \mathbf{a} + s \cdot \mathbf{g} + \mathbf{e}_1 \pmod Q$, $\mathbf{v} = -s \cdot \mathbf{u} - \gamma\cdot \mathbf{g} + \mathbf{e}_2 \pmod Q$.
     Set \( b=\mathbf{b}[0] \) and \( a=\mathbf{a}[0] \) which are the first components of \( \mathbf{b} \) and \(\mathbf{a} \), respectively.
       The subscripts is used to identify keys associated with distinct key holders.

  \item $ct\leftarrow\texttt{CDKS.Encrypt}(\mu,pk)$:
  Samples \( w \leftarrow \chi \) and \( e_0, e_1 \leftarrow D_\sigma \).
  \textcolor{blue}{Encrypts the given plaintext $\mu\in R$ and outputs the ciphertext $ct=w\cdot pk+(\mu+e_0,e_1)\pmod Q$.}
  \textcolor{red}{(Encrypts the given plaintext $\mu\in R_t$ and returns the ciphertext $ct=w\cdot pk+(\lfloor(Q/t)\cdot \mu\rceil+e_0,e_1)\pmod Q$).}

  \item $\overline{ct}:=\texttt{CDKS.Expand}\left(\{pk_i\}_{i\in[1,n]};j;ct\right)$: Expands the given ciphertext $ct$, encrypted under $pk_j$, into a expanded ciphertext $\overline{ct}$ associated with $n$ public keys $\{pk_i\}_{i\in[1,n]}$. Specifically, a ciphertext $ct=(c_0,c_1)$ is expanded into the ciphertext $\overline{ct}=(\bar{c}_0,\cdots,\bar{c}_n)\in R_Q^{n+1}$, where $\bar{c}_0=c_0,\bar{c}_j=c_1$, and the remaining $\{\bar{c}_i\}_{i\in[1,n],i\neq \{0,j\}}$ are all set to 0.

  \item $\overline{ct}_{\text{add}}\leftarrow\texttt{CDKS.Add}(\overline{ct},\overline{ct}')$:
  For the given ciphertexts $\overline{ct},$ $\overline{ct}'\in $ $R_{Q}^{n+1}$, the addition is performed as \( \overline{ct}_{\text{add}} = \overline{ct}+\overline{ct}' \pmod Q \).

  \item $\overline{ct}_{\text{mult}}\leftarrow\texttt{CDKS.Mult}(\overline{ct},\overline{ct}'; \{evk_i\}_{i\in[1,n]} )$:
  For the given ciphertexts $\overline{ct}=(c_i)_{i\in[0,n]},\overline{ct}'= (c_i')_{i\in[0,n]}\in R_{Q}^{n+1}$ and their associated evaluation keys \( \{evk_i\}_{i\in[1,n]} \),
  the multiplication is performed as \textcolor{blue}{\( ct_{\text{mult}} = (c_{i}c_j')_{i,j\in[0,n]}\pmod Q \)} \textcolor{red}{($ct_{\text{mult}} = (\lfloor(t/Q) c_i \cdot c_j' \rceil)_{i,j\in[0,n]}\pmod Q$)}.
  Run Algorithm~\ref{alg:RelinCDKS} with \( (ct_{\text{mul}}, \{evk_i\}_{i\in[1,n]}) \) and output the result $\overline{ct}_{\text{mult}}$.

  \item $\mu\leftarrow\texttt{CDKS.Decrypt}(\overline{ct},\overline{sk})$:  Decrypts a given ciphertext $\overline{ct}=(\bar{c}_0,\bar{c}_1,\cdots,\bar{c}_n)$ using $\overline{sk}=(1,s_1,\cdots,s_n)$ to obtain $\mu=\langle\overline{sk},\overline{ct}\rangle\pmod Q$.
    \textcolor{blue}{Return $\mu$} \textcolor{red}{(Return $\mu=\lfloor(t/Q)\cdot\mu\rceil$)}.
\end{itemize}

\begin{algorithm}[!h]
\caption{CDKS Relinearization}
\begin{algorithmic}[1]
\Require $ct_{\text{mult}} = (c_{i,j})_{i,j\in[0,n]} \in R_{Q}^{(n+1) \times (n+1)}$, \par
$\{evk_i=(\mathbf{b}_i, \mathbf{d}_i, \mathbf{u}_i,\mathbf{v}_i)\}_{i\in[1,n]}$
\Ensure $\overline{ct}_{\text{mult}} = (\bar{c}_i)_{i\in[0,n]} \in R_{Q}^{n+1}$
\State $\bar{c}_0 \gets c_{0}c_0'$
\For{$i = 1$ to $n$}
    \State $\bar{c}_i \gets c_{0}c_i' + c_{i}c_0' \pmod{Q}$
\EndFor
\For{$i = 1$ to $n$}
    \For{$j = 1$ to $n$}
        \State $\bar{c}_j \gets \bar{c}_j + c_{i}c_j' \boxdot \mathbf{d}_i \pmod{Q}$
        \State $\bar{c}_i \gets \bar{c}_i + c_{i}c_j' \boxdot \mathbf{b}_j \pmod{Q}$
        \State $(\bar{c}_0, \bar{c}_i) \gets (\bar{c}_0, \bar{c}_i) + c_{i}c_j' \boxdot (\mathbf{v}_i, \mathbf{u}_i) \pmod{Q}$
    \EndFor
\EndFor
\end{algorithmic}
\label{alg:RelinCDKS}
\end{algorithm}

\noindent In CDKS, the distributed decryption process is as follows: 
\begin{itemize}[leftmargin=10pt,itemsep=0pt, topsep=0pt, parsep=0pt, partopsep=0pt]
  \item $\nu_{i}\leftarrow\texttt{CDKS.PartDec}(\overline{c}_i,s_{i})$:  Partially decrypts the given ciphertext component $\overline{c}_i$ using $s_{i}$ and outputs $\nu_{i}=\overline{c}_is_{i}+e_i\pmod Q$, where $e_i\leftarrow D_{\sigma}$. 

  \item $\mu:=\texttt{CDKS.Merge}(\overline{c}_0,\{\nu_{i}\}_{i\in [1,n]})$:
  Merges the results of partial decryption to compute $\mu=\bar{c}_0+\sum_{i=1}^n \nu_i\pmod {Q}$. \textcolor{blue}{Return $\mu$} \textcolor{red}{(Return $\mu=\lfloor(t/Q)\cdot\mu\rceil$)}.
\end{itemize}
 The distributed decryption process is correct since
\begin{equation*}
      \mu=\textcolor{red}{\lfloor(t/Q)\cdot}\big(\bar{c}_0+\sum_{i=1}^n \nu_i \big) \textcolor{red}{\rceil} 
      \approx \textcolor{red}{\lfloor(t/Q)\cdot}\langle\overline{sk},\overline{ct}\rangle\textcolor{red}{\rceil}\pmod Q.
\end{equation*}

\subsection{Security Vulnerability of CDKS}\label{ssec:securityVulneraCDKS}

\begin{figure}
 \centering
 \includegraphics[width=2.3in]{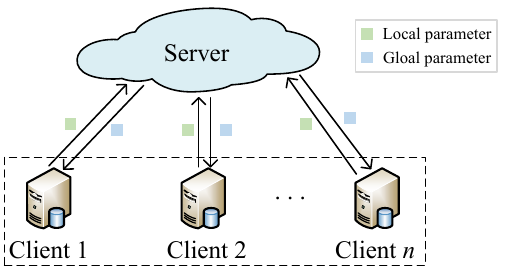}
 \caption{FL system model.}
 \label{fig:FLsystem}
\end{figure}

The CDKS scheme fails to provide the level of security it claims. 
As a result, existing MPC applications built upon CDKS, such as secure FL \cite{tifs2024application} and secure distributed sparse gaussian process \cite{aaai2024application}, inherit CDKS's inherent vulnerabilities and consequently fail to achieve their intended security guarantees.
This limitation arises from a fundamental flaw in CDKS that allows either the server or the clients to recover plaintexts. 
To illustrate this issue, consider an FL system with $n$ clients and an aggregation server (as depicted in Fig.~\ref{fig:FLsystem}) secured using CDKS.
The core process unfolds as follows:
\begin{itemize}[leftmargin=10pt,itemsep=0pt, topsep=0pt, parsep=0pt, partopsep=0pt]
  \item \textbf{Key generation:} Each client $i$ generates its secret-public key pair $(sk_i,pk_i)$.

  \item \textbf{Encryption:} Each client $i$ trains its local model and encrypts its local paremeter $\mu_i$ into a fresh ciphertext $ct_i=(c_0^i, c_1^i)$. Then, it sends $ct_i$ to the server.

  \item \textbf{Secure aggregation of local updates:} Upon receiving $ct_i=(c_0^i, c_1^i)$, the server applies \texttt{CDKS.Expand} to expand the ciphertext into
  $\overline{ct}_i=(c_0^i,0,\cdots,0, c_1^i,0,\cdots,0)$, where $c_0^i$ and $c_1^i$ are replaced in the first and $(i+1)$th positions, respectively, while the remaining positions are padded with zeros.
  Then, by using \texttt{CDKS.Add}, the server aggregates all expanded ciphertexts $\{\overline{ct}_i\}_{i\in[1,n]}$ into the global ciphertext
  $$\overline{ct}=\big(\sum_{i=1}^n c_0^i, c_1^1,c_1^2,\cdots,c_1^n\big)\overset{\triangle}{=}(\bar{c}_0,\bar{c}_1,\bar{c}_2,\cdots,\bar{c}_n),$$
  which is then sent to all clients for decryption.

  \item \textbf{Decryption to obtain a global update}:
  Each client $i$ first computes its partial decryption result $\nu_i\leftarrow\texttt{CDKS.PartD}$- $\texttt{ec}(\bar{c}_i,s_i)=c_1^i s_i + e_i\pmod Q (e_i\leftarrow D_{\sigma}),$
   and broadcasts $\nu_i$ to the other clients or sends it to the aggregation server.
   Then, any client or the server computes the final decryption result as the global update:
   {\small
   \begin{align*}
      &\mu:= \texttt{CDKS.Merge}(\bar{c}_0,\{\nu_i\}_{i\in[1,n]})
      = \textcolor{red}{\lfloor(t/Q)\cdot}\big(\bar{c}_0+\sum_{i=1}^n \nu_i\big) \textcolor{red}{\rceil}\nonumber\\
     &= \textcolor{red}{\lfloor(t/Q)\cdot}\big(\sum_{i=1}^n (c_0^i+c_1^i\cdot s_i)+\sum_{i=1}^ne_i\big) \textcolor{red}{\rceil}\approx \sum_{i=1}^n \mu_i\ (\text{mod}\ Q)
    \end{align*}
    }
\end{itemize}

Although the above process ensures correct parameter aggregation, it exposes local client parameters even under the honest-but-curious assumption, where both the server and the clients are assumed to faithfully execute the FL protocol while being curious to infer the privacy data of other clients. 
This security vulnerability stems from the ciphertext expansion and distributed decryption processes.
Specifically, each component $\bar{c}_i$  ($i\in[1,n]$) of the expanded ciphertext $\overline{ct}=(\bar{c}_0,\bar{c}_1,\bar{c}_2,\cdots,\bar{c}_n)$ is associated with only the fresh ciphertext $c_1^i$ of the client $i$.
This design facilitates decryption correctness, but
introduces a critical security vulnerability.
Since $c_0^i$ is publicly available -- either directly from the fresh ciphertext $ct_i = (c_0^i, c_1^i)$ or from the expanded ciphertext $\overline{ct}_i$, and the partial decryption result $\nu_i$ is shared with the server or the clients for full decryption, any party with access to $\nu_i$ can trivially recover client $i$'s plaintext by computing $\textcolor{red}{\lfloor(t/Q)\cdot}\big(c_0^i + \nu_i \big) \textcolor{red}{\rceil}$. That is,
\begin{equation}\label{eq:inference}
\textcolor{red}{\lfloor(t/Q)\cdot}\big(c_0^i + \nu_i \big) \textcolor{red}{\rceil}=\textcolor{red}{\lfloor(t/Q)\cdot}\big(c_0^i+s_ic_1^i \big) \textcolor{red}{\rceil}\approx\mu_i\pmod Q,
\end{equation}
revealing the client $i$'s original plaintext $\mu_i$.

This compromise demonstrates that CDKS fails to provide adequate security guarantees in MPC scenarios.
To address the security limitations of CDKS,
in this work, we propose secure MKHE schemes, named SMHE, to achieve secure multi-key variants of CKKS and BFV.

\section{New Multi-Key Variants of CKKS and BFV}\label{sec:SMHE}

To achieve our SMHE scheme, we retain the foundational structure of CDKS but introduce a masking scheme. 
This masking scheme ensures both the security and correctness of homomorphic addition, thereby enabling secure aggregation applications such as FL.
Furthermore, ciphertexts processed through homomorphic addition with the masking scheme can still be directly evaluated by the homomorphic multiplication operations of CDKS without additional requirements or auxiliary computations.

\subsection{A Masking Scheme for CKKS and BFV}\label{ssec:maskingScheme}

We design a masking scheme tailored for CKKS and BFV, highlighting its role as a critical component in MKHE framework. Essentially, a masking scheme enables the use of a CKKS/BFV public key $pk$ (with an associated secret key $sk=(1,s)$) and a plaintext $\mu$ to generate a triple $(ct,cz,\bm{\Gamma})$. Here, $ct=(c_0,c_1)$ represents CKKS/BFV encryption of $\mu$ under $pk$, while $cz=(z_0,z_1)$ and $\bm{\Gamma}=(\bm{\varsigma}_0,\bm{\varsigma}_1)$ act as auxiliary information with two key properties: (1) the triple $(ct,cz,\bm{\Gamma})$ provides computational hiding for $\mu$, similar to $ct$ alone, and (2) when provided with another CKKS/BFV public key $pk'$ (corresponding to a secret key $sk'=(1,s')$), it becomes feasible to construct a pair of polynomials $cx=(x_0,x_1)\in R_Q^2$ according to $\bm{\Gamma}$, such that
$\langle sk, cx\rangle+\langle sk', cz\rangle\approx 0$ and
$\textcolor{red}{\lfloor(t/Q)\cdot}\big(\langle sk,ct\rangle+\langle sk, cx\rangle+\langle sk', cz \rangle\big) \textcolor{red}{\rceil}\approx \mu\pmod Q$.
The latter also implies that $\textcolor{red}{\lfloor(t/Q)\cdot}\langle(1,s,s'),(c_0+x_0+z_0,c_1+x_1,z_1)\rangle\textcolor{red}{\rceil}=\textcolor{red}{\lfloor(t/Q)\cdot}\langle \overline{sk},\overline{ct} \rangle \textcolor{red}{\rceil}\approx \mu\pmod Q$. Here, $\overline{sk}=(1,s,s')$ is the concatenated secret key and $\overline{ct}=(c_0+x_0+z_0,c_1+x_1,z_1)$ is a masked expanded ciphertext.

\textbf{CKKS/BFV Masking Scheme.}  The masking scheme contains a triple of algorithms defined as follows:
\begin{itemize}[leftmargin=10pt,itemsep=0pt, topsep=0pt, parsep=0pt, partopsep=0pt]
  \item $\texttt{UniEnc}\left(\mu, pk\right)$: Given a message $\mu\in R_Q$ and a CKKS/BFV public key $pk$, 
  returns a ciphertext $ct$.
  \item $\texttt{MaskEnc}\left(r,pk\right):$ Given a random masking $r$ and a public $pk$, the masking encryption algorithm outputs a pair $(cz,\bm{\Gamma})$.
  \item $\texttt{Extend}\left(\bm{\Gamma}, pk, pk'\right)$: Provided with $\bm{\Gamma}$ and two CKKS/BFV public keys $pk, pk'$, 
  outputs $cx \in R_Q^2$.
\end{itemize}

\noindent The masking scheme meets the following properties:

Semantic Security:   For a given security parameter $\lambda$,
  the security of CKKS/BFV encryption ensures that:
  \begin{align*}
    (pp, pk, \texttt{UniEnc}(\mu, pk)) &\approx_c (pp, pk, \texttt{UniEnc}(\mu', pk)), \\
    (pp, pk, \texttt{MaskEnc}(r, pk))& \approx_c (pp, pk, \texttt{MaskEnc}(r', pk)),
  \end{align*}
  where $pp\gets \texttt{FHE.Setup}(1^\lambda) $, $ (sk, pk) \gets \texttt{FHE.Keygen}$ $(pp) $, $r'\leftarrow\chi$, and $\mu'\leftarrow R_Q$.
  $\texttt{FHE}$ represents the traditional single-key CKKS/BFV scheme.

Correctness:
  Let $pp\gets \texttt{FHE.Setup}(1^\lambda) $, and consider two independently generated key pairs $(sk, pk) $ and $ (sk', pk') $, obtained from
  $ \texttt{FHE.Keygen}(pp)$. For any $\mu\in R_Q$, let $ct \gets \texttt{UniEnc}(\mu, pk)$, $(cz,\bm{\Gamma}) \gets \texttt{MaskEnc}(r, pk) $,  and $cx \gets \texttt{Extend}(\bm{\Gamma}, pk, pk'). $
  Then
  $\mu := \texttt{FHE.Decrypt}(sk, ct)$ and
  $$ \langle sk,ct\rangle+\langle sk, cx\rangle+\langle sk', cz\rangle= \mu + e,$$
  where $ \|e\|_\infty \leq (2N^2 + 4N)B_\chi+\tau N\cdot B_\chi B_{\mathcal{H}} $.

\vspace{5pt}

\textbf{Instantiation.}
We now instantiate our masking scheme.

\begin{itemize}[leftmargin=10pt,itemsep=5pt, topsep=0pt, parsep=0pt, partopsep=0pt]
  \item $\texttt{UniEnc}(\mu, pk):$
  Given a plaintext $\mu\in R_Q$ and a public key $pk=(b,a)\in R_Q^2$, sample $w\leftarrow\chi$ and $e_{w_0},e_{w_1}\leftarrow D_{\sigma}$.
   Encrypts $\mu$ with $pk$ using CKKS/BFV encryption algorithm to output a fresh ciphertext $ct$:
      \begin{align}\label{eq:encryption}
          ct &\leftarrow \texttt{FHE.Encrypt}(pk, \mu)\\
          & =w\cdot pk + (\textcolor{red}{\lfloor(Q/t)\cdot}\mu\textcolor{red}{\rceil}+e_{w_0},e_{w_1}) \pmod Q. \nonumber
        \end{align}

  \item $\texttt{MaskEnc}(r,pk):$  Given a random masking $r\in R_Q$, this algorithm outputs the masking ciphertexts $(cz,\bm{\Gamma})$:

    \begin{enumerate}[leftmargin=12pt,itemsep=2pt, topsep=0pt, parsep=0pt, partopsep=0pt]
    \item 
    Sample $e_{r_0},e_{r_1}\leftarrow D_{\sigma}$ and encrypt the value $0$ with $r$ and $pk$ using CKKS/BFV encryption algorithm and output
      \begin{equation}\label{eq:encryption0}
        \begin{split}
          cz &\leftarrow \texttt{FHE.Encrypt}(pk, 0; r)\\
          & = r\cdot pk+ (e_{r_0},e_{r_1}) \pmod Q. \\
        \end{split}
      \end{equation}

      \item 
          Perform gadget encryption on $r$ and output $\bm{\Gamma}\leftarrow\texttt{GgtEnc}(sk,r)$, which meets $\langle sk,\bm{\Gamma}\rangle\approx r\cdot \mathbf{g}\pmod Q$.
  \end{enumerate}

  \item $\texttt{Extend}(\bm{\Gamma}, pk, pk')$: On input $\bm{\Gamma}\in R_Q^{\tau\times 2}$ and public keys $pk=(b,a)\in R_Q^2,pk'=(b',a)\in R_Q^2$, the algorithm outputs
  $cx=(b'-b)\boxdot \bm{\Gamma}$, such that $\langle sk,cx\rangle\approx r(b'-b)$.

  \item $\texttt{Extend}^*(\bm{\Gamma}, pk,\{pk_{i}\}_{i\in [1,n]})$: This algorithm takes $\bm{\Gamma}\in R_Q^{\tau\times 2}$ and the public keys $pk$ and $\{pk_{i}\}_{i\in [1,n]}$ as input, computes $\sum_{i\in [1,n]}(b_{i}-b)=\sum_{i\in [1,n]}b_{i}-nb$, and outputs $cx=\left(\sum_{i\in [1,n]}(b_{i}-b)\right)\boxdot \bm{\Gamma}$, such that $\langle sk,cx\rangle\approx r\cdot\sum_{i\in [1,n]}(b_{i}-b)$.
\end{itemize}

\noindent The $\texttt{Extend}^*$ algorithm simplifies the process of running $\texttt{Extend}$ for all $n$ public key $\{pk_{i}\}_{i\in [1,n]}$.
Rather than executing $\texttt{Extend}(\bm{\Gamma}, pk, pk_{1})+\cdots+\texttt{Extend}(\bm{\Gamma}, pk, pk_{n})=\sum_{i\in [1,n]}(b_{i}-b)\boxdot\bm{\Gamma}$, it consolidates these computations into a single execution of $\texttt{Extend}^*(\bm{\Gamma}, pk,\{pk_{i}\}_{i\in [1,n]})$.

\textbf{Semantic Security.}
The attacker's view consists of the distribution $\left(pp, pk, ct,cz, \bm{\Gamma}\right)$,
where $pp \gets \texttt{SMHE.Setup}(1^\lambda)$,
$ (sk, pk) \gets \texttt{SMHE.}\texttt{Keygen}(pp)$,
$ct \gets \texttt{UniEnc}(\mu,pk) $,
$(cz,$ $\bm{\Gamma}) \gets \texttt{MaskEnc}(r,pk),\mu\in R_Q$.
The semantic security proof leverages the security of CKKS/BFV encryption and follows these steps:
(1) Modify $\bm{\Gamma}$ to the gadget encryption of random polynomial instead of the gadget encryption of $r$. 
(2) Replace $ct$ and $cz$ with encryptions of random messages. 
These are justified by the semantic security of CKKS/BFV encryption, as the random sample $w$ is unknown and the random sample $r$ is no longer accessible after encryption.
Thus, the distribution becomes independent of $\mu$, establishing semantic security.

\textbf{Correctness.}
Let $\{(sk, pk), (sk' , pk')\}$ represent two valid key pairs produced by $\texttt{SMHE.Keygen}(pp)$.
Recall that $sk =(1,s) ,sk' =(1,s'); s,s'\leftarrow \chi; pk = (b,a) \in R_q^2, pk' = (b',a) \in R_q^2$ with $ b = -a\cdot s + e\pmod Q, b' = -a\cdot s' + e' \pmod Q$, and $ \|e\|_\infty, \|e'\|_\infty \leq \beta_\chi$.

The masking ciphertext with public key $pk$ is $(cz,\bm{\Gamma}) \gets \texttt{MaskEnc}(r,pk)$.
For a message $\mu$, let $ct \gets \texttt{UniEnc}(\mu, pk)$ and $cx \gets \texttt{Extend}(\bm{\Gamma}, pk, pk') $. 
We have
  \begin{align}\label{eq:decryptionCKKS}
    \langle sk, ct\rangle &=\langle (1,s),(wb+\textcolor{red}{\lfloor(Q/t)\cdot}\mu\textcolor{red}{\rceil}+e_{w_0},wa+e_{w_1})\rangle \\
      & =\textcolor{red}{\lfloor(Q/t)\cdot}\mu\textcolor{red}{\rceil}+we+e_{w_0}+se_{w_1}       =\textcolor{red}{\lfloor(Q/t)\cdot}\mu\textcolor{red}{\rceil}+ e_c,\nonumber\\
    \langle sk', cz\rangle & =\langle (1,s'),(rb+e_{r_0},ra+e_{r_1})\rangle \\
      & =rb+e_{r_0}+ras'+s'e_{r_1}
       =r(b - b') + e_c',\nonumber
  \end{align}
where $e_c=we+e_{w_0}+se_{w_1}$  and $e_c'=re'+e_{r_0}+s'e_{r_1}$ with $\|e_c\|_\infty,\|e_c'\|_\infty \leq (N^2 +2N)B_\chi $.
Due to the correctness of linear combinations, we can also deduce that
\begin{equation}\label{eq:masking2}
  \begin{split}
     \langle sk, cx\rangle&  = \langle sk,(b'-b)\boxdot (\bm{\varsigma}_0,\bm{\varsigma}_1)\rangle \\
      & =\langle\mathcal{H}(b'-b),\langle sk,(\bm{\varsigma}_0,\bm{\varsigma}_1)\rangle\rangle=r(b'-b)+e_r ,
  \end{split}
\end{equation}
where $e_r=\langle\mathcal{H}(b'-b),\mathbf{e}\rangle$ and $ \|e_r\|_\infty \leq \tau N\cdot B_\chi B_{\mathcal{H}}$.
Combining these results, we can obtain that $\langle sk, cx\rangle+\langle sk', cz\rangle\approx 0$ and
$\textcolor{red}{\lfloor(t/Q)\cdot}\big(\langle sk, ct\rangle+\langle sk, cx\rangle+\langle sk', cz\rangle \big) \textcolor{red}{\rceil}
=\textcolor{red}{\lfloor(t/Q)\cdot}\textcolor{red}{\lfloor(Q/t)\cdot}\mu\textcolor{red}{\rceil}\textcolor{red}{\rceil}+e^*\approx\mu\pmod Q,$
where $ \|e^*\|_\infty \leq (2N^2 + 4N)B_\chi+\tau N\cdot B_\chi B_{\mathcal{H}} $, as required.

\begin{algorithm}[!htbp]
\small
\caption{SMHE addition algorithm}
\begin{algorithmic}[1]
\Require 
$\overline{ct} = (c_i)_{i\in[0,n]}$ and $\overline{ct}' = (c'_i)_{i\in[0,n]}$, associated with public keys $\{pk_j\}_{j\in T}$ and $\{pk_j\}_{j\in T'}$, respectively; 
The involved masking parameters $\{cz_k,\bm{\Gamma}_k\}_{k\in T\cup T'}$.
\Ensure $\overline{ct}_{add} = (\bar{c}_i)_{i\in[0,n]} \in R_{Q}^{n+1}$
\State $\bar{c}_0=c_0+c_0'\pmod{Q}$
\State $\{cx_i=(0,0)\}_{i\in [1,n]}$, $\{cz_i'=(0,0)\}_{i\in [1,n]}$
\For{$i\in[1,n]$}
    \If{$i \in T$}
        \State $cx_i\leftarrow \texttt{Extend}^*(\bm{\Gamma}_i,pk_i,\{pk_j\}_{j\in T'\setminus i}) \pmod{Q}$
        \State $cz_i'\leftarrow \sum_{j\in  T'\setminus i}cz_j \pmod{Q}$
    \EndIf
    \If{$i \in T'$}
        \State {\small$cx_i\leftarrow cx_i+\texttt{Extend}^*(\bm{\Gamma}_i,pk_i,\{pk_j\}_{j\in T\setminus i}) \pmod{Q}$}
        \State {\small$cz_i'\leftarrow cz_i'+\sum_{j\in  T\setminus i}cz_j \pmod{Q}$}
    \EndIf
    \State $(\bar{c}_0,\bar{c}_i)\gets (\bar{c}_0,c_i+c_i' )+cx_i+cz_i' \pmod{Q}$
\EndFor
\end{algorithmic}\label{alg:mkheAdd3}
\end{algorithm}

\subsection{SMHE Construction}
We construct our SMHE based on the proposed masking scheme, which contains the following algorithms:

\begin{itemize}[leftmargin=10pt,itemsep=2pt, topsep=0pt, parsep=0pt, partopsep=0pt]
  \item[$\bullet$] $pp\leftarrow\texttt{SMHE.Setup}(1^{\lambda})\rightarrow pp$:  Takes as input a security parameter $\lambda$ and outputs the system parameters $pp=\{N,\textcolor{red}{t},Q,\chi\,\sigma,\mathbf{a},\mathbf{\mathcal{H}},\mathbf{g}\}$, where $N=N(\lambda)$ is the RLWE dimension; \textcolor{red}{$t\in \mathbb{Z}$ is the plaintext modulus}; $Q=\sum_{i=1}^L q_i$ is the ciphertext modulus; $\chi$ and $D_{\sigma}$ are the key distribution and error distribution, respectively; $\mathbf{a}\leftarrow U(R_q^d)$ is a CRS; $\mathbf{\mathcal{H}}:R_Q\rightarrow R^{\tau}$ and $\mathbf{g}\in R_Q^{\tau}$ are a gadget decomposition and its gadget vector, respctively.

  \item[$\bullet$] $(sk,pk,evk)\leftarrow\texttt{SMHE.KeyGen}(pp)$: Generates a secret-public key pair $(sk=(1,s),pk=(b,a))$ and an evaluation key $evk=(\mathbf{b},\mathbf{d},\mathbf{u},\mathbf{v})$ following the CDKS construction.

  \item[$\bullet$] \(C\leftarrow\texttt{SMHE.Encrypt}(\mu, pk)\):
  This algorithm encrypts the given plaintext \textcolor{blue}{\(\mu\in R\)} \textcolor{red}{($\mu\in R_t$)} into the ciphertext $ct$ and generates masking parameters $\{cx,\bm{\Gamma}\}$ under the encryption key $pk$ by using the functions \(\texttt{UniEnc}(\mu, pk)\) and $\texttt{MaskEnc}(r,pk)$, respectively. Specifically, the algorithm samples \( w,r \leftarrow \chi \) and \( e_{v0}, e_{v1},e_{r0},e_{r1} \leftarrow D_\sigma \) and computes \textcolor{blue}{$ct=w\cdot pk+(\mu+e_{v0},e_{v1})\pmod Q$}
  \textcolor{red}{($ct=w\cdot pk+(\lfloor(Q/t)\cdot \mu\rceil+e_{v0},e_{v1})\pmod Q$)},
  $cz=r\cdot pk+ (e_{r_0},e_{r_1}))\pmod Q$
  and $\bm{\Gamma}\leftarrow\texttt{GgtEnc}(sk,r)$. 
  Finally, the algorithm outputs the tuple $C=\{ct,cx,\bm{\Gamma}\}$.

  \item[$\bullet$]\(\overline{ct}:=\texttt{SMHE.Expand}(\{pk_{i}\}_{i\in[1,n]},j,ct) \):
    The algorithm expands the given ciphertext \( ct = (c_0, c_1) \), encrypted under \( pk_j \), into an expanded ciphertext \( \overline{ct} = (\bar{c}_i)_{i\in[0,n]} \), where:
    \begin{equation*}\label{eq:expandedCiphertext}
        \bar{c}_0 = c_0 \quad \text{and} \quad
        \bar{c}_i=
        \begin{cases}
            c_1 & \text{if } i = j, \\
            0 & \text{otherwise},
        \end{cases}
         \quad \text{for} \ i\in[1,n].
    \end{equation*}
    We denote the associated key and the reference set of the expanded ciphertext $\overline{ct}$ as $pk_j$ and $\{pk_{i}\}_{i\in[1,n]}$, respectively.

  \item[$\bullet$] $\overline{ct}_{\text{add}} \leftarrow \texttt{SMHE.Add}_2\big( \overline{ct}_1, \overline{ct}_2; pk_1,pk_2;\{cz_i, $ $\bm{\Gamma}_i\}_{i \in [1,2]} \big)$:
    This algorithm takes as inputs two expanded ciphertexts $\overline{ct}_1 = (c_0^1,c_1^1,0),\overline{ct}_2 = (c_0^2,0,c_1^2)$ which are obtained by invoking $\texttt{SMHE.Expand}(\{pk_{i}\}_{i\in[1,2]},j,ct_j)$ for $j\in[1,2]$.
    The associated masking parameters of the two parties are $\{cz_i,  \bm{\Gamma}_i\}_{i \in [1,2]}$.
    The algorithm first performs $cx_1=(x_0^1,x_1^1)\leftarrow\texttt{Extend}(\bm{\Gamma}_1, pk_1, pk_2),cx_2=(x_0^2,x_1^2)\leftarrow\texttt{Extend}(\bm{\Gamma}_2, pk_2, pk_1)$.
    Then, it masks the two ciphertexts as
    $\widehat{ct}_1 = (c_0^1+x_0^1+z_0^2,c_1^1,z_0^1), \widehat{ct}_2 = (c_0^2+x_0^2+z_0^1,z_0^2,c_1^2)$\footnotemark
    \footnotetext{
    The masked ciphertexts $\widehat{ct}_1,\widehat{ct}_2$ can be equivalently interpreted as being obtained via an expansion of the masked fresh ciphertexts.
    That is, although the actual implementation performs masking after ciphertext expansion, 
    $\widehat{ct}_1$ and $\widehat{ct}_2$ can be conceptually understood as the expansion of masked ciphertexts: 
    $\widehat{ct}_1 =\texttt{SMHE.Expand}(\{pk_{i}\}_{i\in[1,2]},1,ct_1+cx_1+cz_2),$
    $\widehat{ct}_2 =\texttt{SMHE.Expand}(\{pk_{i}\}_{i\in[1,2]},2,ct_2+cx_2+cz_1).$
    This conceptual view clarifies the design intuition behind our masking strategy: masking is applied before expansion to hide plaintexts while enabling multi-key compatibility.
    }.
    Finally, the algorithm outputs $\overline{ct}_{\text{add}} = \widehat{ct}_1+\widehat{ct}_2\pmod Q$.

  \item[$\bullet$] $\overline{ct}_{\text{add}} \leftarrow \texttt{SMHE.Add}\big( \overline{ct}, \overline{ct}'; \{pk_j\}_{j \in T},\{pk_j'\}_{j \in T'};\{cz_k,$ $\bm{\Gamma}_k\}_{k \in T \cup T'} \big)$:
    This algorithm is a general case of homomorphic addition.
    It takes as inputs two ciphertexts $\overline{ct} = (c_i)_{i \in [0,n]},\overline{ct}' = (c'_i)_{i \in [0,n]} \in R_q^{n+1}$, the corresponding reference sets $\{pk_j\}_{j \in T}$ and $\{pk_j'\}_{j \in T'}$, and the involved masking parameters $\{cz_k,$ $\bm{\Gamma}_k\}_{k \in T \cup T'}$.
    It invokes Algorithm~\ref{alg:mkheAdd3} 
    on the inputs and outputs the homomorphic addition result \(\overline{ct}_{\text{add}} = (\bar{c}_i)_{i \in [0,n]}\).

  \item[$\bullet$] $\overline{ct}_{\text{mult}}\leftarrow\texttt{SMHE.Mult}(\overline{ct},\overline{ct}'; \{evk_i\}_{i\in[1,n]} )$:
  For two given ciphertexts $\overline{ct}=(c_i)_{i\in[0,n]},\overline{ct}'= (c_i')_{i\in[0,n]}\in R_Q^{n+1}$ and their associated public keys \( \{pk_i\}_{i\in[1,n]} \),
  the multiplication is performed as \textcolor{blue}{\( \overline{ct}_{\text{mult}} = (c_{i}c_j')_{i,j\in[0,n]}\ (\text{mod } Q) \)} \textcolor{red}{($\overline{ct}_{\text{mult}} = (\lfloor(t/Q) c_i \cdot c_j' \rceil)_{i,j\in[0,n]}\ (\text{mod } Q)$)}.
  Run Algorithm~\ref{alg:RelinCDKS} with \( (ct_{\text{mul}}, \{evk_i\}_{i\in[1,n]}) \) and output the result $\overline{ct}_{\text{mult}}$.

  \item[$\bullet$] $\nu_{i}\leftarrow\texttt{SMHE.PartDec}(\bar{c}_i,s_{i})$:  Partially decrypts a given ciphertext component $\bar{c}_i$ using $s_{i}$ and outputs $\nu_{i}=\bar{c}_is_{i}+e_i\ (\text{mod } Q)$, where $e_i\leftarrow\chi$.

  \item[$\bullet$] $\mu:=\texttt{SMHE.Merge}(\bar{c}_0,\{\nu_{i}\}_{i\in [1,n]})$:
  Compute $\mu=\bar{c}_0+\sum_{i=1}^n \nu_i\ (\text{mod}\ Q)$. \textcolor{blue}{Return $\mu$} \textcolor{red}{(Return $\mu=\lfloor(t/Q)\cdot\mu\rceil$)}.
\end{itemize}

\textbf{Remark:}
$\texttt{SMHE.Add}_2$ and $\texttt{SMHE.Add}$ both denote the homomorphic addition of two given ciphertexts.
The key difference lies in the underlying computation setting: $\texttt{SMHE.Add}_2$ corresponds to the two-party setting, where the reference set includes only the union of the two parties' public keys; whereas $\texttt{SMHE.Add}$ is defined in the general multi-party setting, where the reference set comprises the union of public keys from all multiple participating parties.

\vspace{-3pt}
\subsubsection{Security of SMHE}\label{sssec:securityAndcorrectness}

We now present the security analysis of the proposed SMHE scheme under the semi-honest adversarial model using a simulation-based proof.

\textbf{Security Setting.}
We consider a semi-honest adversary $\mathcal{A}$ who follows the protocol honestly but may try to infer additional information from observed messages.
To prove security, we follow the real/ideal world simulation paradigm as defined in Definition~\ref{definition:simulationSecurity}, requiring that any view generated in the real-world execution is computationally indistinguishable from a simulated view generated in the ideal world, where only the final output is known.

\textbf{Simulator Construction.}
Consider a set of $n$ parties ($n \geq 2$), each holding a message $\mu_i$ and a key pair $(pk_i, sk_i)$ generated via \texttt{MKHE.KeyGen}.
Let $\mathcal{C}$ be the function evaluated over the encrypted messages, and let $\mu = \mathcal{C}(\mu_1, \dots, \mu_n)$ be the final output.
We construct a PPT simulator $\mathsf{Sim}$ that, given only the public parameters $pp$ and the final output $\mu$, generates a view that is computationally indistinguishable from the adversary's real-world view:
\begin{itemize}[leftmargin=10pt,itemsep=2pt, topsep=0pt, parsep=0pt, partopsep=0pt]
  \item \emph{Encryption Simulation:}
  SMHE employs RLWE-based encryption (e.g., CKKS or BFV) to generate ciphertexts (including masking parameters). Under the RLWE assumption, the ciphertexts are semantically secure. Hence, $\mathsf{Sim}$ can simulate each ciphertext $ct_i$ as a uniformly random element from the ciphertext space, without knowing its plaintext $\mu_i$.

  \item \emph{Evaluation Simulation:}
  The evaluation phase includes ciphertext expansion, masking, and arithmetic operations.
  Due to the introduction of masking terms, the evaluated ciphertext $\overline{ct}$ is statistically independent of the inputs.
  Thus, $\mathsf{Sim}$ can simulate $\overline{ct}$ and intermediate ciphertexts using fresh samples of the appropriate algebraic structure.

  \item \emph{Partial Decryption Simulation:}
  Each party generates a partial decryption share $\nu_i$, which includes fresh RLWE-style noise.
  The simulator $\mathsf{Sim}$ can generate these as random RLWE-like elements that are indistinguishable from actual decryptions.
  The final decryption result $\mu$ is known, so $\mathsf{Sim}$ can simulate the full decryption outcome consistently.
\end{itemize}

\textbf{Security Guarantee.}
Let $\text{View}_{\mathcal{A}}^{\text{real}}$ denote the adversary's view in the real execution, and let $\text{View}_{\mathsf{Sim}}^{\text{ideal}}$ be the simulated view.
Under the RLWE assumption, we have $\text{View}_{\mathcal{A}}^{\text{real}}(\lambda) \approx_c \text{View}_{\mathsf{Sim}}^{\text{ideal}}(\lambda).$
That is, no efficient adversary can distinguish between the real and ideal views with non-negligible probability.
This proves that SMHE achieves simulation-based security against semi-honest adversaries.

\textbf{Remark.}
Although our analysis is in the semi-honest setting, the SMHE scheme can be extended to handle malicious adversaries by incorporating verifiable computation techniques, such as zero-knowledge proofs~\cite{sun2024zkllm} or interactive oracle proofs~\cite{aranha2025heliopolis}, which allow each party to prove correctness of operations without revealing private data.
We leave the integration of such mechanisms to future work.

\subsubsection{Correctness of Homomorphic Evaluation}
We now present the correctness of the proposed homomorphic evaluation algorithm.

\textbf{Correctness of Two-Party Homomorphic Addition ($\texttt{MKHE.Add}_2$).} The correctness of SMHE homomorphic addition is ensured by the correctness of the underlying masking scheme.
  Assume the two parties are $P_1$ and $P_2$, who independently generate their key materials as
  $(sk_i,pk_i,evk_i)\leftarrow\texttt{SMHE.KeyGen}(pp)$ for $i=1,2$.
  Each party $P_i$ encrypts its plaintext $\mu_i$ as $ct_i=(c_0^i,c_1^i)$ and generates its masking parameters $(cz_i,\bm{\Gamma}_i)$ by invoking $\{ct_i,cx_i,\bm{\Gamma}_i\}\leftarrow\texttt{SMHE.Encrypt}(\mu_i, pk_i)$.
  Then, $ct_1,ct_2$ are expanded
   as $$\overline{ct}_1:=\texttt{SMHE.Expand}(\{pk_1,pk_2\},1,ct_1)=(c_0^1,c_1^1,0),$$ $$\overline{ct}_2:=\texttt{SMHE.Expand}(\{pk_1,pk_2\},2,ct_2)=(c_0^2,0,c_1^2),$$ which are further masked and added to obtain
   \begin{align*}
      & \overline{ct}_{add}=(\bar{c}_0,\bar{c}_1,\bar{c}_2)\\
      \leftarrow\ & \texttt{SMHE.Add}_2\left( \overline{ct}_1, \overline{ct}_2; pk_1, pk_2; \{cz_i, \bm{\Gamma}_i\}_{i \in [1,2]} \right) \\
         =\ & (c_0^1+c_0^2+x_0^1+x_0^2+z_0^1+z_0^2, c_1^1+x_1^1+z_1^2, c_1^2+x_1^2+z_1^1)
     \end{align*}
    where the masking components are generated as
    $$cz_1=(z_0^1, z_1^1),cx_1=(x_0^1,x_1^1)\leftarrow \texttt{Extend}(\bm{\Gamma}_1, pk_1, pk_2),$$
    $$cz_2=(z_0^2,z_1^2),cx_2=(x_0^2,x_1^2)\leftarrow \texttt{Extend}(\bm{\Gamma}_2, pk_2, pk_1).$$
   Decryption of $\overline{ct}_{add}$ is performed as follows:\\
   Each party $P_i$ computes a partial decryption $\nu_{i}\leftarrow\texttt{SMHE.PartDec}(\bar{c}_i,s_{i})$ $=\bar{c}_is_{i}+e_i \pmod Q,$
   where $e_i\leftarrow\chi$.
   Then, the full decryption result is
   {\small
     \begin{align}\label{eq:CorrectAdd2Party}
        \mu\ & := \texttt{SMHE.Merge}(\bar{c}_0,\{\nu_{i}\}_{i\in [1,n]})\nonumber \\
        =\ & \textcolor{red}{\lfloor(t/Q)\cdot}\big(\bar{c}_0+\bar{c}_1s_{1}+\bar{c}_2s_{2}+e_1+e_2 \big) \textcolor{red}{\rceil}\nonumber \\
        \approx\ & \textcolor{red}{\lfloor(t/Q)\cdot}\langle(1,s_1,s_2), (c_0^1+c_0^2+x_0^1+z_0^2+x_0^2+z_0^1, \\
        &\ c_1^1+x_1^1+z_1^2, c_1^2+x_1^2+z_1^1)\rangle\textcolor{red}{\rceil}\nonumber \\
        =\ & \textcolor{red}{\lfloor(t/Q)\cdot}\big(\langle sk_1,ct_1+cx_1+cz_2\rangle+\langle sk_2,ct_2+cx_2+cz_1\rangle \big) \textcolor{red}{\rceil}   \nonumber \\
       \approx \  &\textcolor{red}{\lfloor(t/Q)\cdot}\big(\langle sk_1,ct_1\rangle+\langle sk_2,ct_2\rangle \big) \textcolor{red}{\rceil}    \approx \mu_1+\mu_2 \pmod Q. \nonumber
     \end{align}}
  Eq.~\eqref{eq:CorrectAdd2Party} holds due to the correctness of the masking scheme, which ensures that
  $\langle sk_1,(cx_1+cz_2)\rangle+\langle sk_2,(cx_2+cz_1)\rangle\ (\text{mod } Q) \approx 0$.
  Therefore, the correctness of the homomorphic addition for two parties in SMHE is achieved.

\textbf{Correctness of Generalized Multi-Party Homomorphic Addition ($\texttt{MKHE.Add}$).}
  Let $\overline{ct}=(c_i)_{i\in[0,n]}$ and $\overline{ct}'=(c_i')_{i\in[0,n]}$ represent two expanded ciphertexts associated with encryption keys $\{pk_j\}_{j\in T}$ and $\{pk_j\}_{j\in T'}$, respectively.
  Denote the plaintexts corresponding to these ciphertexts are $\mu$ and $\mu'$, respectively.
  That is, $\textcolor{red}{\lfloor(t/Q)\cdot}\langle \overline{sk},\overline{ct}\rangle \textcolor{red}{\rceil}\approx \mu \pmod Q$ and
  $\textcolor{red}{\lfloor(t/Q)\cdot}\langle \overline{sk},\overline{ct}'\rangle \textcolor{red}{\rceil}\approx \mu' \pmod Q$ hold,
  where $\overline{sk}=(1,s_1,\cdots,s_n)$.
  The masking parameters associated with the two ciphertexts are given as $\left\{\left(cz_k,\bm{\Gamma}_k\right)\right\}_{k\in T\cup T'}$.
  The result of the homomorphic addition between the two ciphertexts is $\overline{ct}_{add}=(\bar{c}_k)_{k\in[0,n]}\leftarrow\texttt{SMHE.Add}\big( \overline{ct}, \overline{ct}'; \{pk_j\}_{j \in T}, \{pk_j\}_{j \in T'}; \{cz_k, \bm{\Gamma}_k\}_{k \in T \cup T'} \big)$ and its decryption proceeds as follows:
{\small
\begin{align}\label{eq:CorrectAdd}
   &\quad\textcolor{red}{\lfloor(t/Q)\cdot}\langle \overline{sk},\overline{ct}_{add}\rangle \textcolor{red}{\rceil}   =  \bar{c}_0 + \sum_{i\in [1,n]}s_i\cdot \bar{c}_i \nonumber \\
      & = \textcolor{red}{\lfloor(t/Q)\cdot} \big((c_0+c_0')+\sum_{i\in [1,n]}s_i\cdot (c_i+c_i')\big)  \textcolor{red}{\rceil} \\
      &\quad +\textcolor{red}{\lfloor(t/Q)\cdot}\sum_{i\in T\cup T'}\left\langle sk_i, \left(cx_i+cz_i'\right)\right\rangle \textcolor{red}{\rceil}  \nonumber  \\
      & \approx \textcolor{red}{\lfloor(t/Q)\cdot}\big(\langle \overline{sk},\overline{ct}\rangle + \langle \overline{sk},\overline{ct}'\rangle \big) \textcolor{red}{\rceil}
      \approx \mu+\mu'\pmod {Q}.\nonumber
\end{align}}
\noindent Eq.~\eqref{eq:CorrectAdd} holds due to the correctness of the masking scheme, which ensures that
  $\sum_{i\in T\cup T'}\left\langle sk_i, \left(cx_i+cz_i'\right)\right\rangle\ (\text{mod}\ Q)\approx 0$.
  Therefore, the correctness of the homomorphic addition in SMHE is achieved.

\textbf{Correctness of Homomorphic Multiplication in a General Case  ($\texttt{MKHE.Mult}$).}
  The correctness of homomorphic multiplication in SMHE is ensured by the correctness of ciphertext multiplications proposed in \cite{ccs2019mkhe} and \cite{ccs2023mkhe}. While this applies to the multiplication of two expanded ciphertexts derived from fresh ciphertexts, we now demonstrate that the correctness also holds for multiplications between an expanded ciphertext and a masked ciphertext, as well as for multiplications between two masked ciphertexts.
\begin{itemize}[leftmargin=10pt,itemsep=2pt, topsep=0pt, parsep=0pt, partopsep=0pt]
  \item Correctness of multiplications between an expansion ciphertext and a masked ciphertext:
            Let $\widehat{ct}=(\hat{c}_i)_{i\in[0,n]}$ denote a masked ciphertext,
            for example, $\widehat{ct}=\texttt{SMHE.Add}\big(\overline{ct},\overline{ct}'; $ $ \{pk_j\}_{j \in T}, \{pk_j\}_{j \in T'}; \{cz_k, \bm{\Gamma}_k\}_{k \in T \cup T'}\big)$, with a plaintext value of $\mu+\mu'$. Here, $\overline{ct}=(c_i)_{i\in[0,n]}$ and $\overline{ct}'=(c_i')_{i\in[0,n]}$ are two expanded ciphertexts defined as earlier.
            The result of the homomorphic multiplication between the masked ciphertext $\widehat{ct}$ and the expanded ciphertext $\overline{ct}'$ is  $\overline{ct}_{\text{mult}}=$ $(\bar{c}_i)_{i\in[0,n]}\leftarrow\texttt{SMHE.Mult}(\widehat{ct},\overline{ct}'; \{evk_i\}_{i\in[1,n]})$, where
            {\small
            \begin{align*}
                \bar{c}_0=\ &  \textcolor{red}{\lfloor(t/Q)\cdot}(
                            \hat{c}_0\cdot c_0'+\!\!\sum_{i\in[1,n]}(\!\!\sum_{j\in[1,n]}\!\!\hat{c}_i\cdot c_j'\boxdot \mathbf{b}_j)\boxdot \mathbf{v}_i
                            ) \textcolor{red}{\rceil} (\text{mod}\ {Q}), \\
                 \bar{c}_i=\ & \textcolor{red}{\lfloor(t/Q)\cdot}\big(\hat{c}_i\cdot c_0'+\hat{c}_0\cdot c_i'+\sum_{j\in[1,n]}\hat{c}_j\cdot c_i'\boxdot \mathbf{d}_j \\
                        &+\big(\sum_{j\in[1,n]}\hat{c}_i\cdot c_j'\boxdot \mathbf{b}_j\big)\boxdot \mathbf{u}_i\big) \textcolor{red}{\rceil} \pmod {Q}
              \end{align*}  }      for $i\in[1,n]$.
            The decryption of $\overline{ct}_{\text{mult}}$ proceeds as follows:
            {\small
              \begin{align*}
                &\quad \textcolor{red}{\lfloor(t/Q)\cdot} \langle \overline{sk}, \overline{ct}_{\text{mult}}\rangle  \textcolor{red}{\rceil}
                =\textcolor{red}{\lfloor(t/Q)\cdot}\big(\bar{c}_0 + \sum_{i\in[1,n]}s_i\cdot \bar{c}_i\big) \textcolor{red}{\rceil}\\
                &=\textcolor{red}{\big\lfloor(t/Q)\cdot \lfloor(t/Q)\cdot}\big(\hat{c}_0 c_0'+\sum_{i\in[1,n]}s_i(\hat{c}_i\cdot c_0'+\hat{c}_0\cdot c_i') \\
                &+\sum_{i,j=1}^n(\hat{c}_i c_j'\boxdot \mathbf{d}_i)\cdot s_j
                 + \hat{c}_i c_j'\boxdot \mathbf{b}_j(\mathbf{v}_i+s_i\cdot \mathbf{u}_i) \textcolor{red}{\rceil\big\rceil}\!\!\!\!  \pmod Q.
              \end{align*}}
            According to the relinearization algorithm in \cite{ccs2019mkhe,ccs2023mkhe}, the following approximation holds:
            $$(\hat{c}_i c_j'\boxdot \mathbf{d}_i)\cdot s_j + \hat{c}_i c_j'\boxdot \mathbf{b}_j(\mathbf{v}_i+s_i\cdot \mathbf{u}_i)\approx\hat{c}_i c_j'\cdot s_i s_j\  (\text{mod}\ {Q}).$$
            Thus, the decryption simplifies to (Set $s_0=1$):
            {\small
            \begin{align*}
                & \textcolor{red}{\lfloor(t/Q)\cdot} \langle \overline{sk}, \overline{ct}_{\text{mult}}\rangle  \textcolor{red}{\rceil}
                     \approx  \textcolor{red}{\big\lfloor(t/Q)\cdot \lfloor(t/Q)\cdot}\!\!\!\sum_{i\in[0,n]}\sum_{j\in[0,n]}\!\!\hat{c}_ic_j'\cdot s_is_j\textcolor{red}{\rceil\big\rceil}\\
                  & = \textcolor{red}{\lfloor(t/Q)\cdot}\langle \overline{sk}, \widehat{ct}\rangle \textcolor{red}{\rceil}
                      \! \cdot\! \textcolor{red}{\lfloor(t/Q)\cdot}\langle \overline{sk}, \overline{ct}'\rangle \textcolor{red}{\rceil}=(\mu+\mu')\cdot\mu'\!\!\!\!  \pmod Q.
              \end{align*}}
            Therefore, the correctness of the homomorphic multiplication between a masked ciphertext and an expanded ciphertext is established.

      \item Correctness of multiplications between two masked ciphertexts: Similarly, the correctness of the homomorphic multiplication between two masked ciphertexts can be demonstrated by applying the same reasoning and leveraging the properties of the relinearization algorithm.
  \end{itemize}

\begin{figure*}[!htbp]
\centering
  \includegraphics[width=5in]{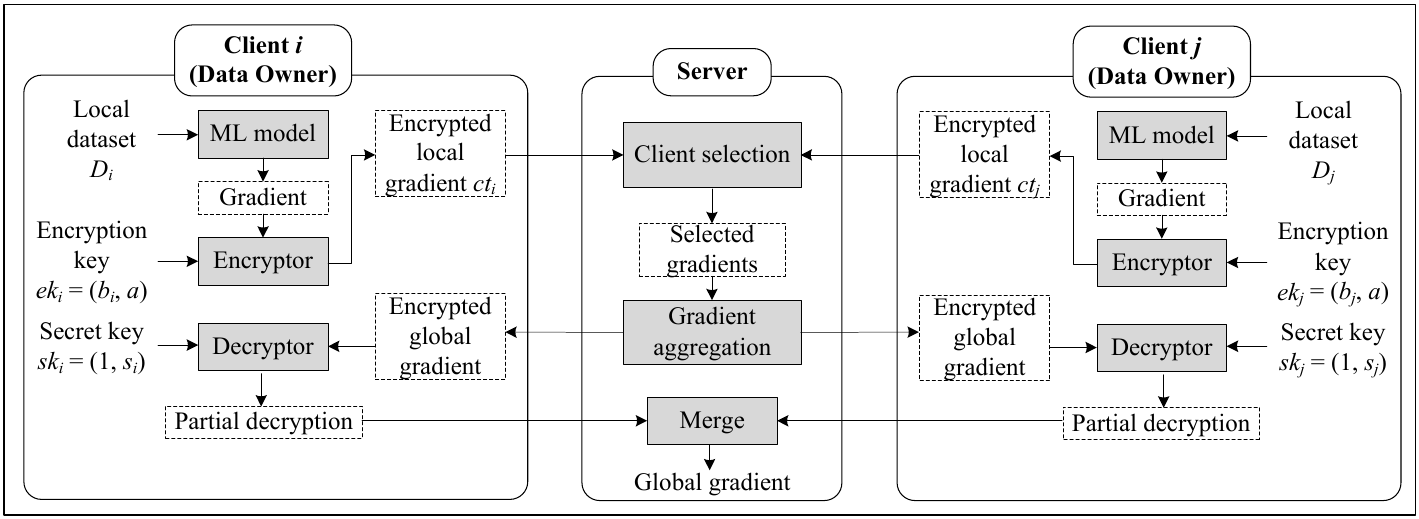}
 \caption{\label{fig:overview} High-level overview of privacy-preserving federated learning.}
\end{figure*}

\section{ Application to Privacy-Preserving FL}\label{sec:wholeModel}
We use the proposed SMHE to construct the PPFL model, as illustrated in Fig.~\ref{fig:overview}.
Next, we introduce the threat model and outline the specific phases of our PPFL model.

\subsection{Threat Model}

We adopt a threat model similar to the one presented in the CDKS schemes \cite{ccs2019mkhe,ccs2023mkhe} and their privacy-preserving distributed computing applications \cite{tifs2024application,aaai2024application}. The threat model assumes the presence of multiple participants, including clients and the server, each with specific roles and assumptions regarding their behavior.

\subsubsection{Participants and Assumptions}
\begin{itemize}
  \item \textbf{Clients:} Each client in the federated learning system holds private data and performs local training on their dataset. We assume that clients are honest-but-curious: they faithfully follow the protocol and participate in the FL process. However, they may be curious about the data and model updates from other clients, potentially attempting to infer private information.

  \item \textbf{Server:} The server is responsible for aggregating model updates received from clients and generating the global model. Similar to clients, the server is assumed to be honest-but-curious, faithfully executing the protocol but potentially interested in learning about the data and updates from individual clients.
\end{itemize}

\subsubsection{Privacy and Security Goals}
The main objective of the threat model is to ensure that sensitive information remains protected during the FL process. Specifically, the goal is to prevent the leakage of client data to other participants. The model must ensure that no participant can learn anything about others' private data beyond what is revealed by the output of the computation. This aligns with the security objectives of MPC, where input privacy is a fundamental goal, ensuring that private data remains confidential even during collaborative computations.

\subsection{Specific Phase of the PPFL Model}
As depicted in Fig.~\ref{fig:overview}, the PPFL model involves the following steps:

\noindent\textbf{Initialization}: Public parameters and keys generation.
    \begin{itemize}[leftmargin=10pt,itemsep=2pt, topsep=0pt, parsep=0pt, partopsep=0pt]
      \item \textit{Public parameter generation.} The server and the clients negotiate the security parameter $\lambda$ and generates the public parameters $pp=\{N,\textcolor{red}{t},Q,\chi\,\sigma,\mathbf{a},\mathbf{\mathcal{H}},g\}\leftarrow\texttt{SMHE.Setup}(1^{\lambda})$ an initial global model parameter $\hat{w}^0$.
      \item \textit{Key generation.}
          Each client $i$ generates its key pair $(sk_i=(1,s_i),pk_i=\{b_i,a\})$ by executing $\texttt{SMHE.KeyGen}(pp)$.
    \end{itemize}
\vspace{5pt}

\noindent\textbf{Model training and encryption (Client side)}: Each client $i$ locally trains its model and encrypts its local update.
    \begin{itemize}[leftmargin=10pt,itemsep=2pt, topsep=0pt, parsep=0pt, partopsep=0pt]
      \item \textit{Model initialization.} Initializes the local model as $w_i^t\leftarrow \hat{w}^{(t-1)}$, where $t\geq 1$ represents the $t$-th round of training.

      \item \textit{Training.}
        Trains its local model $w_i^t$ with a mini-batch dataset to generate a local gradient $g_i^t$ (a.k.a local model update).

      \item \textit{Encryption.} 
          The client encodes $g_i^t$ as $\mu_i\in R$, encrypts $\mu_i$ as $\{ct_i,cz_i,\bm{\Gamma}_i\}\leftarrow\texttt{SMHE.Encrypt}(\mu_i,ek_i)$,
          and sends $\{ct_i,cz_i,\bm{\Gamma}_i\}$ to the server.
      \end{itemize}
\vspace{5pt}

\noindent\textbf{Secure aggregation (Server side)}: After receiving all the $n$ clients' submissions, the server performs secure aggregation.
    \begin{itemize}[leftmargin=10pt,itemsep=2pt, topsep=0pt, parsep=0pt, partopsep=0pt]
      \item \textit{Ciphertext expandsion.}
       The server expands the ciphertexts $ct_{i}=(c_{i,0},c_{i,1})$ $({i}\in [1,n])$ into \( \overline{ct}_{i} = (\bar{c}_{i,j})_{j\in[1,n]} \) by invoking $\texttt{SMHE.Expand}(\{pk_{j}\}_{j\in[1,n]}; i; ct_i) $. Here,
            \[
            \bar{c}_{i,0} = c_{i,0} \quad \text{and} \quad
            \bar{c}_{i,j}=
            \begin{cases}
                c_{i,1} & \text{if } j = i, \\
                0 & \text{otherwise},
            \end{cases}
            \ \text{for}\ j\in[1,n]
            \]

        \item \textit{Weighted aggregation.} The server selects some clients and aggregates their local gradients as a global gradient.
        Denote the selected client set as $\mathcal{P}_{benign}$ with the size $m=|\mathcal{P}_{benign}|$.
        The server aggregates all the $m$ ciphertexts $\{\overline{ct}_{id_i}\}_{{id_i}\in \mathcal{P}_{benign}}$ as an aggregation ciphertext $\overline{ct}_{add}^m$.
        Specifically, the server computes
        \begin{align*}
            \overline{ct}_{add}^1&\leftarrow\texttt{SMHE.Add}(\alpha_{id_1}\overline{ct}_{id_1},(0)_{m+1};pk_{id_1},pk_{id_m};\\
            & \{cz_{id_1},  \bm{\Gamma}_{id_1}\}, \{cz_{id_m},\bm{\Gamma}_{id_m}\}), \ \text{and}\\
            \overline{ct}_{add}^{i}&\leftarrow\texttt{SMHE.Add}(\alpha_{id_{i-1}}\overline{ct}_{add}^{i-1},\alpha_{id_i}\overline{ct}_{id_i};pk_{id_{i-1}},pk_{id_i}; \\
            &\{cz_{id_{i-1}},  \bm{\Gamma}_{id_{i-1}}\}, \{cz_{id_{i}}, \bm{\Gamma}_{id_i}\})\ \text{for}\ i\in[2,m],
          \end{align*}
where $\alpha_{id_i}\in \mathbb{Q}$ serves as a possible weight in the model.
        Here, we have $id_{i-1}=id_m$ when $i=1$, and $id_{i+1}=id_1$ when $i=m$.
        The aggregation ciphertext is $\overline{ct}_{add}^m\overset{\triangle}{=}(\bar{c}_0,\bar{c}_{id_1},\cdots,\bar{c}_{id_m})$ and $\bar{c}_{id_i}$ is then sent to the client $id_i$. 
    \end{itemize}
\vspace{5pt}

\noindent\textbf{Partial decryption (Client side)}: Each client $id_i\in \mathcal{P}_{benign}$ computes the partial decryption result
$\nu_{id_i}=\texttt{SMHE.PartDec}(\bar{c}_{id_i},$ $s_{id_i})$ and sends $\nu_{id_i}$ to the server.
\vspace{5pt}

\noindent\textbf{Full decryption and model update (Server side)}:
\begin{itemize}[leftmargin=10pt,itemsep=2pt, topsep=0pt, parsep=0pt, partopsep=0pt]

  \item \textit{Full decryption.}
    After receiving $\{\nu_{id_i}\}_{id_i\in \mathcal{P}_{benign}}$ from all the $m$ clients, 
    the server obtains the final decryption result
    $\mu\leftarrow\texttt{MKHE.Merge}(\bar{c}_0,\{\nu_{id_i}\}_{id_i\in \mathcal{P}_{benign}})$.
    Then the server decodes $\mu$ as the weighted aggregation gradient $\hat{g}_t$.

  \item \textit{Parameter update.}
      The server updates the global model as $\hat{w}_{t}:=\hat{w}_{t-1}-\eta \frac{\hat{g}_t}{\sum_{i=1}^m \alpha_i}$ and sends $\hat{w}_{t}$ to the clients for the next iteration until the final training model is obtained.
\end{itemize}


\subsection{Security}\label{ssec:security-ppfl}

We formalize the security of the PPFL protocol instantiated with our structured SMHE from Section \ref{sec:SMHE}. The proof follows the standard \emph{simulation-based} paradigm and relies on the RLWE assumption and the semantic security of CKKS/BFV as well as the masking mechanism in Section \ref{ssec:maskingScheme}.

\textbf{Functionality.}
In one aggregation round there are clients $\{P_i\}_{i=1}^n$ and a server $\mathcal S$.
Client $P_i$ holds a local update $\mu_i$ (e.g., a gradient).
The ideal functionality $\mathcal F_{\mathsf{Agg}}$ outputs only
\[
  \mu \;\approx\; \sum_{i=1}^{n}\mu_i
\]
to $\mathcal S$ (or to all parties, as required by the application), and reveals nothing else about any individual $\mu_i$.

\textbf{Real protocol (recap).}
The real system is the SMHE-based PPFL, where each client encrypts its update under CKKS augmented with a masking triple $(ct,cz,\mathbf{\Gamma})$; the server homomorphically evaluates the additive aggregation circuit; decryption is \emph{distributed}: parties output partial decryptions $\nu_i$ over the reference set, which are linearly merged to recover $\mu$.

\textbf{Adversary and leakage.}
We consider a semi-honest adversary $\mathcal A$ who may corrupt the server and an arbitrary subset $\mathcal P\subseteq \mathcal{P}_{benign}$ of clients while following the protocol. The leakage function $\mathcal L$ includes public parameters, the number of parties, the circuit/model shape, and the functionality output $\mu$.

\begin{definition}[Simulation-based security target]\label{def:sim-target}
Let $\text{View}^{\mathrm{real}}_{\mathcal A}$ be the adversary's view in the real execution
(including $pp,\{pk_i\}$, input/expanded/evaluated ciphertexts, partial decryptions $\{\nu_i\}$, and optionally $\mu$).
Let $\text{View}^{\mathrm{ideal}}_{\mathsf{Sim}}$ be the output of a PPT simulator given only $(pp,\mu)$.
The protocol securely realizes $\mathcal F_{\mathsf{Agg}}$ if
\[
  \text{View}^{\mathrm{real}}_{\mathcal A}(\lambda) \;\approx_c\; \text{View}^{\mathrm{ideal}}_{\mathsf{Sim}}(\lambda).
\]
\end{definition}

\textbf{Why masking matters (intuition).}
A CDKS-style expansion and distributed decryption enables recovering \emph{each} $\mu_i$
from public expansion components and the corresponding partial decryption (cf.\ Eq.~(14) in Section~\ref{sec:priorWork}),
contradicting the ``only-the-sum'' leakage of $\mathcal F_{\mathsf{Agg}}$.
Our SMHE attaches a masking triple $(ct,cz,\mathbf{\Gamma})$ with an auxiliary ciphertext $c_x$ (derived from $\mathbf{\Gamma}$ under a second key) such that
\begin{equation*}
  \begin{split}
    & \langle sk,c_x\rangle + \langle sk',c_z\rangle \approx 0 \pmod Q \\
      & \langle sk,ct\rangle + \langle sk,c_x\rangle + \langle sk',c_z\rangle\approx \mu \pmod Q.
  \end{split}
\end{equation*}
Hence, adversarially observable components are \emph{masked} and become informative only upon the final linear merge.

\begin{theorem}[One-round PPFL security, semi-honest]\label{thm:one-round}
Under the RLWE assumption and the semantic security of CKKS and the masking mechanism in Section~\ref{ssec:maskingScheme}, the SMHE-based PPFL protocol securely realizes $\mathcal F_{\mathsf{Agg}}$ in the semi-honest model:
for any PPT adversary $\mathcal A$ corrupting $\mathcal S$ and any subset $\mathcal P\subseteq \mathcal{P}_{begin}$ of clients, there exists a PPT simulator $\mathsf{Sim}$ such that $\text{View}^{\mathrm{real}}_{\mathcal A}\approx_c \text{View}^{\mathrm{ideal}}_{\mathsf{Sim}}$.
\end{theorem}

\begin{proof}[Proof sketch with hybrids]
We construct $\mathsf{Sim}$ given only $(pp,\mu)$ and proceed via a sequence of hybrids.

\smallskip
\noindent\textbf{H0 (Real execution).}
This is the real protocol with honest clients encrypting their $\mu_i$, the server evaluating, and all parties outputting partial decryptions.

\smallskip
\noindent\textbf{H1 (Simulated honest encryptions/expansions).}
For each honest $i\notin\mathcal P$, replace input and expanded ciphertexts by \emph{fresh, distribution-consistent} RLWE-type samples (same ring/modulus/layout) that are independent of $\mu_i$.
Indistinguishability follows from the semantic security of CKKS and of the masking triple: real encryptions/masks are computationally indistinguishable from such samples.

\smallskip
\noindent\textbf{H2 (Simulated evaluated ciphertexts).}
Replace all intermediate/evaluated ciphertexts derived from honest inputs by fresh RLWE-type samples with the same algebraic shape as the real evaluation.
Because masking ensures that only the \emph{final} linear merge un-masks the data,
any single intermediate component remains computationally independent of the underlying $\mu_i$.
Thus H1 and H2 are indistinguishable.

\smallskip
\noindent\textbf{H3 (Simulated partial decryptions with merge consistency).}
For each honest party, replace $\nu_i$ by an RLWE-type sample drawn from the same distribution as a real partial decryption (inner product plus noise).
To ensure that the final output equals $\mu$, fix one honest index $j$ and set $\tilde{\nu}_j$ \emph{after} sampling the others so that the public linear relation used by the merge procedure satisfies
\[
  Merge\big(\{\tilde{\nu}_i\},\text{public ciphertext components}\big)\approx\mu.
\]
This is always possible because the merge is linear and, by masking correctness,
\[
  \langle sk,ct\rangle + \langle sk,c_x\rangle + \langle sk',c_z\rangle\approx \mu \pmod Q.
\]
Thus H2 and H3 are indistinguishable.

\smallskip
\noindent\textbf{Corrupted parties.}
For every corrupted client $i\in\mathcal P$, $\mathsf{Sim}$ uses the \emph{actual} input and keys (already known to $\mathcal A$) to produce the corresponding values; for the corrupted server, $\mathsf{Sim}$ releases the simulated evaluated ciphertexts and $\{\tilde{\nu}_i\}$ as above.

\smallskip
\noindent\textbf{Conclusion.}
Let $\text{View}^{\mathrm{ideal}}_{\mathsf{Sim}}$ be the distribution in H2.
By transitivity of computational indistinguishability from H0 to H3, we obtain
$\text{View}^{\mathrm{real}}_{\mathcal A}\approx_c \text{View}^{\mathrm{ideal}}_{\mathsf{Sim}}$,
which proves the claim.
\end{proof}

\begin{corollary}[Server + subset collusion]\label{cor:collusion}
Let $\mathcal A$ corrupt the server and any subset $\mathcal P$ of clients.
The protocol remains simulation-secure:
there exists $\mathsf{Sim}$ given $(pp,\mu)$ and $\{\mu_i\}_{i\in\mathcal P}$ (trivially known to $\mathcal A$) that outputs an indistinguishable view by instantiating real values for corrupted parties and applying the hybrids above to honest parties.
\end{corollary}

\textbf{Multi-round composition.}
If each round uses fresh SMHE randomness (fresh encryption noise, fresh $(cz,\mathbf{\Gamma})$, fresh evaluations), then the per-round transcripts are independent; by a standard sequential composition theorem, over $I$ rounds the joint view is indistinguishable from an ideal system that reveals only $\{\mu^{(i)}\}_{i=1}^I$.

\textbf{Client dropout.}
If some clients are absent in a round, the reference set and merge range over the \emph{active} clients only.
Linearity of merge and masking correctness immediately yield that only the sum over active parties is revealed; the same simulator applies unchanged.

\begin{remark}[On inference from functionality leakage]
The ideal functionality $\mathcal F_{\mathsf{Agg}}$ reveals the aggregate $\mu$ only. However, if $\mathcal A$ controls $n-1$ clients, then it can \emph{in the ideal world} derive the remaining honest input by subtraction: $\mu_j=\mu-\sum_{i\in\mathcal P}\mu_i$. Thus preventing per-client inference beyond functionality leakage requires at least two honest clients in the round.
\end{remark}

\begin{remark}[On CDKS vs. SMHE]
CDKS-style expansion plus distributed decryption admits per-client reconstruction (cf.\ Eq.~(14)), hence cannot realize $\mathcal F_{\mathsf{Agg}}$ which leaks only the aggregate.
In contrast, SMHE's masking guarantees that \emph{no pre-merge} component is informative about any individual $\mu_i$, and only the final linear merge reveals $\mu$.
\end{remark}

\subsection{Correctness}\label{ssec:correctness}
We now analyze the correctness of the ciphertext aggregation and decryption processes within PPFL.

During the aggregation process, the aggregated ciphertext is computed as $\overline{ct}_{\text{add}}^1 \leftarrow \texttt{SMHE.Add}(\alpha_{id_1}\overline{ct}_{id_1},(0)_{m+1}; id_1, id_m; $ $ \{cz_{id_1}, \bm{\Gamma}_{id_1}\}, \{cz_{id_m}, \bm{\Gamma}_{id_m}\}),$
yielding:
\begin{align*}
    \overline{ct}_{\text{add}}^1 =\ & (\alpha_{id_1}c_0^{id_1} + x_0^{id_1,id_m}+ z_0^{id_m} + x_0^{id_m,id_1} + z_0^{id_1} , \\
      &\alpha_{id_1}c_1^{id_1} + x_1^{id_1,id_m} + z_1^m, 0, \ldots, 0, x_1^{id_m,id_1} + z_1^{id_1}),
  \end{align*}
and the final aggregated ciphertext is denoted as $\overline{ct}_{\text{add}}^m \overset{\triangle}{=} (\bar{c}_0, \bar{c}_{id_1}, \ldots, \bar{c}_{id_m})$, where
\begin{align*}
  &\bar{c}_0 =\sum_{id_i\in \mathcal{P}_{benign}} \left(\alpha_{id_i}c_0^{id_i} + x_0^{id_{i},id_{i-1}} + x_0^{id_i,id_{i+1}} + 2z_0^{id_i}\right), \\
  &\bar{c}_{id_i} =\alpha_{id_i}c_1^{id_i} + x_1^{id_{i},id_{i-1}} + x_1^{id_i,id_{i+1}} + z_1^{id_{i-1}} + z_1^{id_{i+1}}, 
\end{align*}
for $i\in[1,m]$. Here, 
$cx_{id_i,id_j} = (x_0^{id_i,id_j}, x_1^{id_i,id_j}) \leftarrow \texttt{Extend}(\bm{\Gamma}_{id_i}, pk_{id_i}, pk_{id_j})$ for $id_i,id_j \in \mathcal{P}_{benign}$.

The decryption of the aggregated ciphertext $\overline{ct}_{\text{add}}^n$ is
{\small
\begin{align}\label{eq:decryption}
&\quad \mu = \bar{c}_0 + \sum_{id_i\in \mathcal{P}_{benign}} (\bar{c}_{id_i} \cdot s_{id_i} + e_{id_i}) \\
&= \sum_{id_i\in \mathcal{P}_{benign}} \Big( \alpha_{id_i}\big(c_0^{id_i}+c_1^{id_i} s_{id_i}\big) + e_{id_i} +  \langle sk_{id_i},cx_{{id_i},id_{i-1}} \rangle \nonumber\\
&\quad + \langle sk_{id_i},cz_{id_{i-1}} \rangle + \langle sk_{id_i},cx_{id_i,id_{i+1}} \rangle + \langle sk_{id_i},cz_{id_{i+1}} \rangle \Big) \nonumber\\
&= \alpha_{id_i}\mu_{id_i} + e^* \approx \sum_{id_i\in \mathcal{P}_{benign}} \alpha_{id_i}\mu_{id_i} \pmod Q.\nonumber
\end{align}}

\noindent Eq.~\eqref{eq:decryption} holds due to the correctness of the masking scheme and CKKS' scalar multiplication, which ensures that
\begin{align*}
     & \langle sk_{id_i},cx_{id_i,id_{i-1}} \rangle + \langle sk_{id_i},cz_{id_{i-1}} \rangle\\
      =\ & \langle sk_{id_i},(b_{id_{i-1}}-b_{id_i})\boxdot(\bm{\varsigma}_0^{id_{i-1}},\bm{\varsigma}_1^{id_{i-1}}) \rangle\\
      & \ + \langle sk_{id_i},(r_{id_{i-1}}ek_{id_i}+(e_0,e_1)) \rangle \\
        \approx\ & r_{id_{i-1}}(b_{id_{i-1}}-b_{id_i}) +r_{id_{i-1}}(b_{id_i}-b_{id_{i-1}})= 0\pmod Q,
\end{align*}
and $\langle sk_{id_i}, \alpha_{id_i} ct_{id_i}\rangle\approx\alpha_{id_i}\mu_{id_i}\pmod Q$.
The error bound is 
$\|e^*\|_\infty \leq (3m^3 + 6m^2+m)B_\chi + 2dm^2 \cdot B_\chi B_{\mathcal{H}}.$

\begin{figure*}[!htbp]
  \centering
  \begin{subfigure}[t]{0.48\textwidth}
    \includegraphics[width=\linewidth]{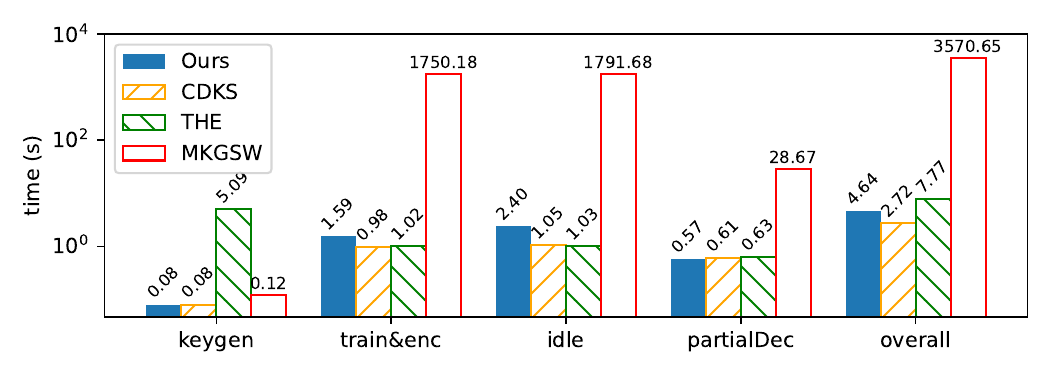}
    \caption{FCN client-side time distribution}
    \label{fig:clientSplitTimeFCN}
  \end{subfigure}
  \begin{subfigure}[t]{0.48\textwidth}
    \includegraphics[width=\linewidth]{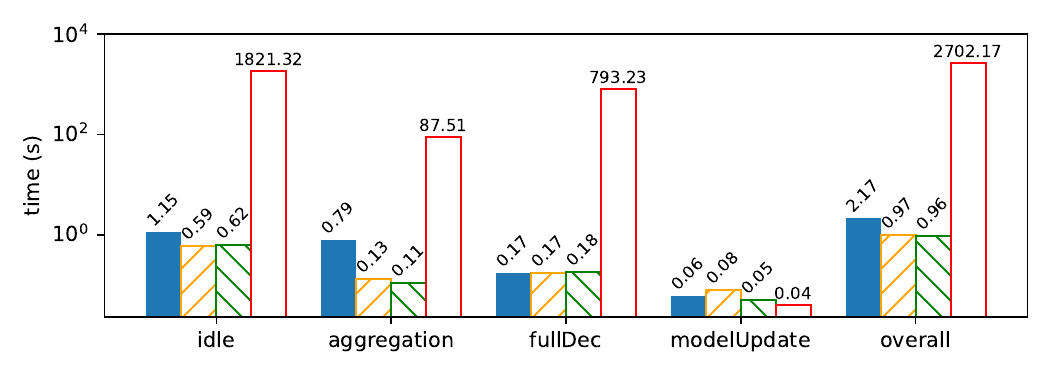}
    \caption{FCN server-side time distribution}
    \label{fig:serverSplitTimeFCN}
  \end{subfigure}
  \hfill
  \begin{subfigure}[t]{0.48\textwidth}
    \includegraphics[width=\linewidth]{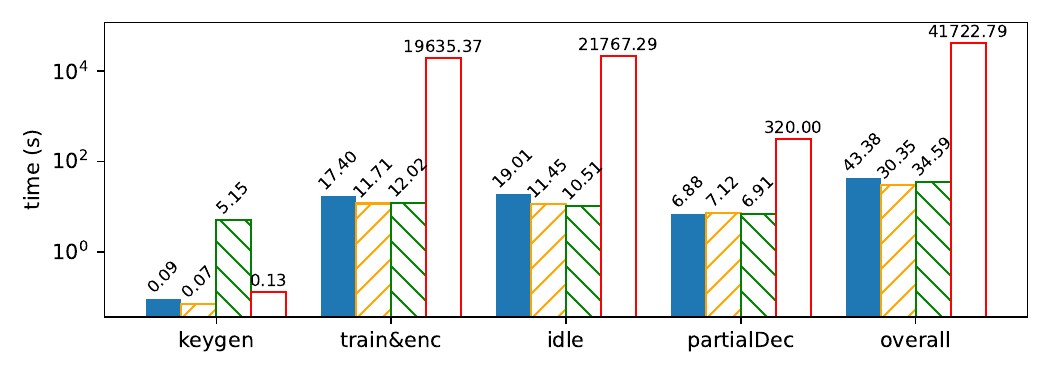}
    \caption{AlexNet client-side time distribution}
    \label{fig:clientSplitTimeAlexNet}
  \end{subfigure}
  \begin{subfigure}[t]{0.48\textwidth}
    \includegraphics[width=\linewidth]{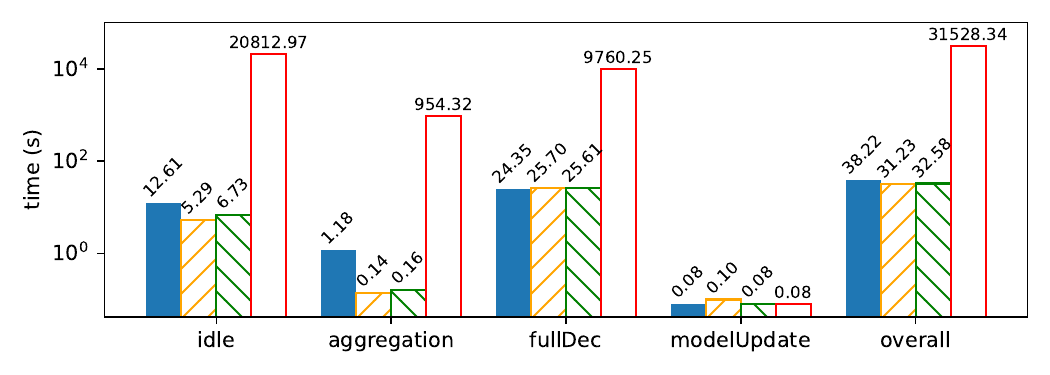}
    \caption{AlexNet server-side time distribution}
    \label{fig:serverSplitTimeAlexNet}
  \end{subfigure}
  \vspace{5pt}
  \caption{Runtime breakdown comparison of PPFL models using different HEs , where ``idle" denotes the duration during which the clients/server remain inactive while waiting for other parties to complete their operations.
  For CDKS and THE, ``train\&enc'' represents the time spent on local model training and encryption. 
  For MKGSW and SMHE, this stage additionally includes the time for masking generation.
  }
  \label{fig:splitTime}
\end{figure*}

\subsection{Implementation}

We conduct PPFL experiments using SMHE, where the multi-key CKKS scheme is applied to encrypt the gradients. 

\subsubsection{Experimental Setup}
The experiments were conducted on a computer equipped with an Nvidia GeForce GTX 1080 Ti GPU and an Intel Core i7-6700 CPU.
The SMHE algorithms were implemented in C++ using the NTL 10.4.0 \cite{ntl} and GMP 6.2.1 \cite{gmp} libraries to handle arbitrary-length integers and high-precision arithmetic.
The FL models were executed using the PyTorch 1.11.0 framework in Python 3.8.
To build our PPFL framework, a dynamic library containing all SMHE-related code was generated and invoked within a Python script.

\textbf{Benchmark Models, Dateset, and Parameter Setting.}
To evaluate the effectiveness of our SMHE-based PPFL model, we conduct comparative experiments against the existing multiparty HE schemes: CDKS \cite{ccs2023mkhe}, THE \cite{ma2022mkhe}, and MKGSW \cite{mukherjee2016two}.
All models are assessed under two distinct learning scenarios to ensure comprehensive and fair comparison:
%
(i) A fully connected neural network (FCN) on the MNIST dataset.
The network includes an input layer with $784$ neurons, two hidden layers with $128$ and $64$ neurons respectively, and an output layer with $10$ neurons representing digit classes from $0$ to $9$ using one-hot encoding.
The model is optimized using the Adam algorithm with a mini-batch size of $64$.
%
(ii) An AlexNet network with approximately $1.25$ million parameters on the CIFAR-10 dataset.
Training is conducted using a batch size of $128$ and the RMSprop optimizer with a decay factor of $10^{-6}$.
%
For both FL tasks, the datasets are randomly partitioned among multiple clients for simulating collaborative training. 

For the SMHE parameters, we largely follow the settings used in CDKS and THE \cite{ccs2019mkhe, ccs2023mkhe, ma2022mkhe} to ensure a fair comparison.
Specifically, the distribution $\chi \leftarrow R$ is instantiated with coefficients drawn from $\{0, 1\}$, and the error term is sampled from a discrete Gaussian distribution $e \leftarrow D_{\sigma}$ with standard deviation $\sigma = 3.2$.
For gadget decomposition, we adopt the RNS-friendly method proposed by Bajard et al. \cite{bajard2016full}. 
We set the gadget vector dimension to $\tau = 8$, the RLWE polynomial dimension to $N = 2^{14}$, the security parameter to $\lambda = 256$, and the number of slots to $n_s = 8192$.
The actual security level under standard RLWE assumptions corresponds to $128$ bits, as estimated by the LWE estimator.
To accelerate the masking encryption and masking extension procedures, we incorporate the SIMD strategy proposed in \cite{wu2025esafl}.
For the MKGSW parameters, we follow the TFHE setting \cite{chillotti2020tfhe} to achieve a $128$-bit security level by setting the LWE dimension to $512$, the modulus to $2^{32}$, and the gadget dimension to $10$.
The computation and communication complexity of SMHE scales linearly with the number of clients $n$, which is consistent with both CDKS and THE, whereas MKGSW exhibits quadratic growth with respect to $n$. Since SMHE inherits the homomorphic multiplication procedure from CDKS without modification, its multiplication complexity remains identical to that of CDKS. Therefore, in our experiments, we focus on comparing the runtime and communication overhead of the homomorphic addition algorithm among SMHE, CDKS, THE, and MKGSW in the context of PPFL applications.

\subsubsection{Experimental Results}
We report the performance of PPFL models using SMHE, CDKS, THE, and MKGSW, focusing on runtime, communication traffic, and model accuracy.

\textbf{Runtime.}
Fig.~\ref{fig:splitTime} presents the runtime breakdown of PPFL models across different computation phases on both the client and server sides for a single training iteration. The results are averaged over $100$ training iterations for all models except the MKGSW-based model, whose runtime is approximated using the MKGSW evaluation time reported in TFHE \cite{chillotti2020tfhe}.
As a bit-level HE scheme, MKGSW incurs an impractically high runtime due to its inherent computational complexity.
For example, under the FCN setting, the total runtime reaches $3570.65$s on the client side and $2702.17$s on the server side.
This excessive overhead renders MKGSW unsuitable for real-world PPFL applications.
In contrast, the other three models, THE-based, CDKS-based, and our SMHE-based models, exhibit significantly lower runtimes, making them more feasible for practical deployment. Among these, THE-based and CDKS-based models demonstrate comparable runtime performance.
Our SMHE-based scheme introduces a moderate increase in runtime relative to THE and CDKS, primarily due to the incorporation of the masking extension mechanism.
For instance, under the AlexNet setting, the total client-side runtime of SMHE increases by less than $2\times$ compared to the CDKS- and THE-based models ($4.64$s vs. $2.72$s and $7.77$s, respectively).
Similarly, on the server side, SMHE incurs a slightly higher overall runtime. 
In all cases, the runtime increase remains within a $2\times$ range, which we consider a reasonable trade-off given the enhanced security guarantees. 

Besides, compared to THE, our approach significantly reduces the client-side key generation time (e.g., $0.08$s vs. $5.09$s for FCN) by allowing each party to generate their key locally, without the need for interactive threshold key generation. This eliminates the communication overhead inherent in THE's distributed key setup, leading to a substantial reduction in key generation and distribution latency.

Overall, SMHE strikes a practical balance between improved security, key flexibility, and runtime efficiency. Compared to MKGSW, it achieves vastly lower computational overhead, while maintaining acceptable runtime increases relative to THE and CDKS. These features make SMHE a scalable and effective solution for MPC applications.

\begin{table}[htbp]
\small
\centering
\caption{Communication traffic comparison. }
 \label{tab:traffic}
\begin{tabular}{|l|c|c|c|}
  \hline
Model\textbackslash{}Network & FCN      & AlexNet  \\ \hline
Plain (\#gradient)           & 109.386K & 1.25M    \\ \hline
THE-based (cipher traffic)   & 7.00MB   & 76.50MB\\  \hline
CDKS-based (cipher traffic)  & 7.00MB   & 76.50MB  \\ \hline
SMHE-based (cipher traffic)  & 16.00MB  & 139.50MB \\ \hline
MKGSW-based (cipher traffic)  & 33.39GB  & 381.47GB \\ \hline
\end{tabular}
\end{table}

\begin{figure}[!htbp]
  \centering
  \begin{subfigure}[t]{0.24\textwidth}
    \centering
    \includegraphics[width=\linewidth]{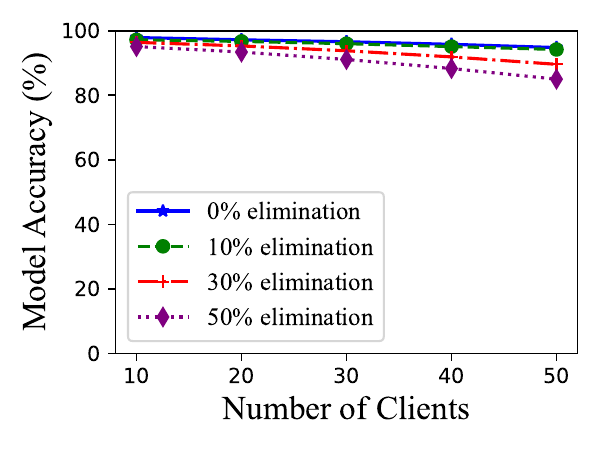}
    \caption{FCN model accuracy for SMHE-based PPFL model}
    \label{fig:mkheFCNAccuracy}
  \end{subfigure}
  \begin{subfigure}[t]{0.24\textwidth}
    \centering
    \includegraphics[width=\linewidth]{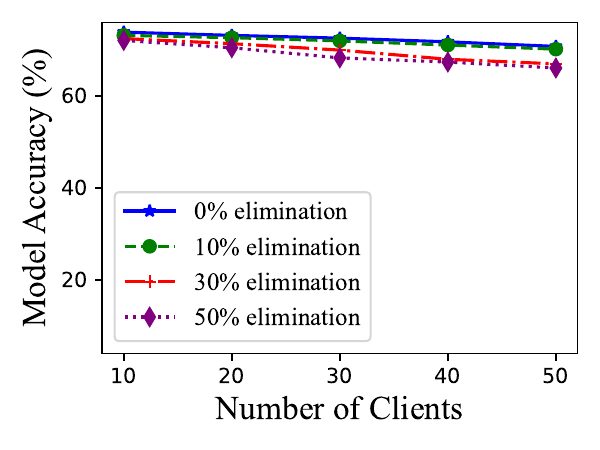}
    \caption{AlexNet model accuracy for SMHE-based PPFL model}
    \label{fig:mkheAlexNetAccuracy}
  \end{subfigure}
  \hfill
  \begin{subfigure}[t]{0.24\textwidth}
    \centering
    \includegraphics[width=\linewidth]{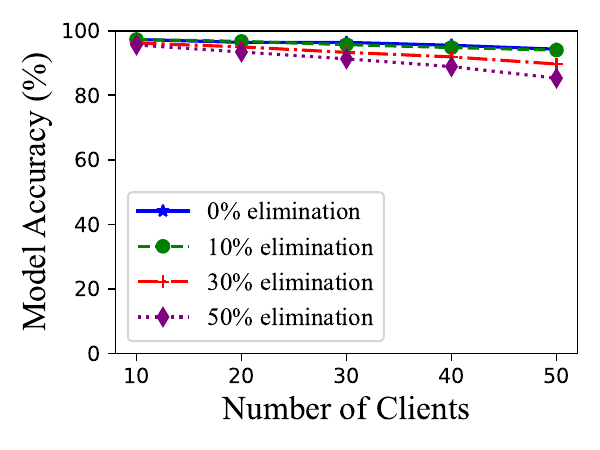}
    \caption{FCN model accuracy for CDKS-based PPFL model}
    \label{fig:cdksFCNAccuracy}
  \end{subfigure}
  \begin{subfigure}[t]{0.24\textwidth}
    \centering
    \includegraphics[width=\linewidth]{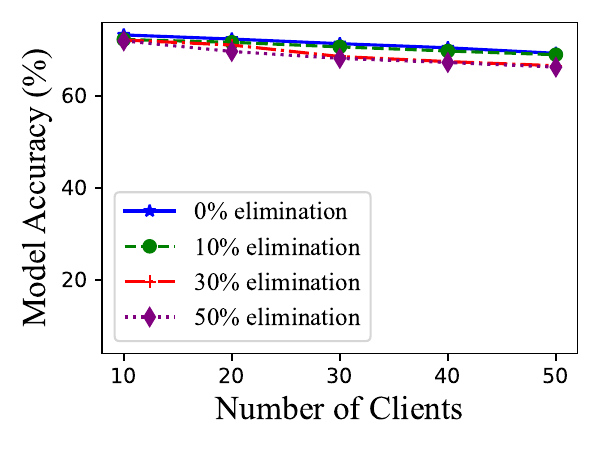}
    \caption{AlexNet model accuracy for CDKS-based PPFL model}
    \label{fig:cdksAlexNetAccuracy}
  \end{subfigure}
  \hfill
  \begin{subfigure}[t]{0.24\textwidth}
    \centering
    \includegraphics[width=\linewidth]{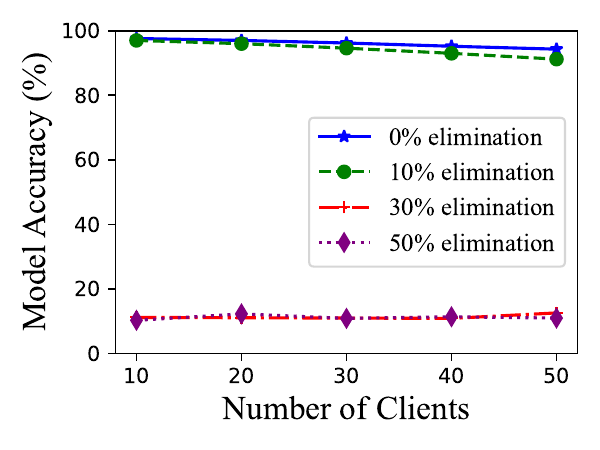}
    \caption{FCN model accuracy for THE-based PPFL model}
    \label{fig:theFCNAccuracy}
  \end{subfigure}
  \begin{subfigure}[t]{0.24\textwidth}
    \centering
    \includegraphics[width=\linewidth]{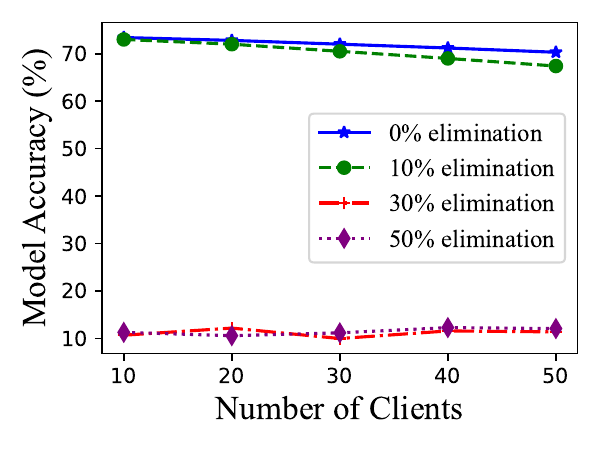}
    \caption{AlexNet model accuracy for THE-based PPFL model}
    \label{fig:theAlexNetAccuracy}
  \end{subfigure}
  \vspace{5pt}
  \caption{Comparison of final model accuracy under different client elimination rates using SMHE/CDKS/THE scheme, where the models are all trained for $10,000$ iterations.}
  \label{fig:accuracy}
\end{figure}

\textbf{Communication Traffic.}
Table~\ref{tab:traffic} presents the communication traffic per iteration for different PPFL models under FCN and AlexNet networks.
The row labeled ``Plain (\#gradient)" reflects the number of model gradients in the plain FL, $109{,}386$K for FCN and $1.25$M for AlexNet.
For encrypted models, the MKGSW-based model introduces an extremely high communication burden, $33.39$GB per iteration for FCN and $381.47$GB for AlexNet, mainly due to its bit-level encryption. This renders MKGSW impractical for real-world PPFL applications, where communication efficiency is a critical concern.
In contrast, the other three models: THE-, CDKS-, and SMHE-based scheme offer significantly lower communication costs and are more suitable for practical deployment.
Specifically, THE- and CDKS-based models share similar ciphertext formats and therefore exhibit identical communication traffic, requiring only $7$MB and $76.5$MB per iteration for FCN and AlexNet, respectively.
Our SMHE-based model incurs a moderate increase in communication traffic, reaching $16$MB for FCN and $139.5$MB for AlexNet. This overhead stems from the ciphertext extension mechanism introduced by our masking design. 
However, despite the increase, the total traffic remains within approximately $2\times$ that of CDKS and THE, which is a reasonable cost given the enhanced security and flexibility.

\textbf{Model Accuracy.} Fig.~\ref{fig:accuracy} displays the final model accuracy of the PPFL models using SMHE, CDKS, and THE
under different client elimination rates.
As observed from the results, the SMHE (CDKS)-based model exhibits strong robustness against client elimination\footnote{While SMHE requires all participants for decryption, its full key independence and dynamism make it particularly suitable for cross-silo FL settings, where participants may join or leave over time but typically remain reliable within each aggregation session.}.
Even with $50$ distributed clients and a high elimination rate of $50\%$, the final model accuracy remains above $85\%$ for FCN and $66\%$ for AlexNet, indicating that the model tolerates a substantial fraction of client drop without significant performance degradation.
In contrast, the THE-based model (where the threshold is set to $0.8$ in the experiment) shows a drastic decline in accuracy once the proportion of eliminated clients exceeds the threshold. The final model accuracy drops sharply, eventually reaching a level close to that of random guessing, indicating the failure of effective model convergence.
These results demonstrate that SMHE-based model offers greater fault tolerance and flexibility in the presence of unreliable or excluded clients, whereas the THE-based system is sensitive to threshold violations due to their dependency on sufficient participant contributions.

\begin{table}[!htbp]
\footnotesize
\centering
\caption{Comparisons of total training time and network traffic for converged FCN and AlexNet networks under different PPFL modes.}
\label{tab:comp_effcy}
\begin{tabular}{|l|l|l|l|l|l|}
\hline
Model                    & Mode       & Epochs & Accuracy & Time(h) & Traffic(GB) \\ \hline
\multirow{4}{*}{FCN}     & Plain       & 34     & 97.94\%   & 0.13      & 0.65      \\ \cline{2-6}
                                         & THE       & 34     & 97.64\%   & 7.75      & 21.85      \\ \cline{2-6}
                                         & CDKS    & 37     & 97.62\%   & 3.57      & 23.77        \\ \cline{2-6}
                                         & SMHE    & 35     & 97.90\%   & 6.22      & 41.40        \\ \hline
\multirow{4}{*}{AlexNet} & Plain      & 268    & 74.01\%   & 4.21      & 48.67     \\ \cline{2-6}
                                         & THE      & 267    & 73.79\%   & 194.29  & 777.92     \\ \cline{2-6}
                                         & CDKS   & 270    & 73.97\%   & 180.12   & 786.67       \\ \cline{2-6}
                                         & SMHE   & 263    & 73.84\%   & 232.49   & 1397.31       \\ \hline
\end{tabular}
\end{table}

\textbf{Training to Convergence.}
We next compare the total training time and network traffic of different PPFL models until convergence is reached.
The number of clients is set to $10$. Due to the exceedingly high time and communication costs involved in the full training process, we simulate federated learning locally until convergence and estimate the total training time and communication cost based on the bandwidth and iteration counts.
Table~\ref{tab:comp_effcy} presents the estimated results for both FCN and AlexNet models, from which we can observe that
%
(1) All PPFL models using SMHE, CDKS, and THE achieve high prediction accuracy, exceeding $97\%$ on FCN and around $74\%$ on AlexNet.
(2) For both the FCN and AlexNet models, SMHE achieves comparable accuracy to the plaintext baseline ($97.90\%$ vs. $97.94\%$ for FCN and $73.84\%$ vs. $74.01\%$ for AlexNet) and
it incurs $6.22$/$232.49$ hours of training time and $41.40$/$1397.31$ GB network traffic for FCN/AlexNet, which are less than $2\times$ that of THE and CDKS.
Overall, compared with THE and CDKS, the SMHE-based PPFL models introduce only a modest increase in training time and communication cost (less than $2\times$) while ensuring stronger security guarantees and offering better dynamism, making it a practical and secure choice for PPFL applications.

\section{Conclusion}\label{sec:conclusion}

In this paper, we propose a novel secure multi-key homomorphic encryption (SMHE) scheme. Our scheme addresses critical security vulnerabilities identified in previous MKHE schemes, specifically those proposed by Chen et al. and Kim et al., which inadvertently reveal plaintext information during multiparty secure computation tasks such as federated learning. By introducing a masking scheme into the CKKS and BFV schemes, our enhanced scheme ensures the confidentiality of data while supporting efficient and secure homomorphic operations.
Future work includes further optimizing the efficiency of ciphertext expansion in SMHE and exploring its integration with advanced applications such as secure model inference and encrypted machine learning.

\bibliographystyle{IEEEtran}
\bibliography{IEEEabrv,cited}



\end{document}